\documentclass[11pt]{article}

\usepackage[dvips]{epsfig}
\usepackage{url}
\newcommand{\solver}[1]{\linebreak[0]{\normalfont\scshape #1}}
\newcommand{\file}[1]{\linebreak[0]{\normalfont\sffamily\small #1}}
\newcommand{\script}[1]{\linebreak[0]{\normalfont\textsl{#1}}}
\newcommand{\command}[1]{{\normalfont\texttt{#1}}}
\newcommand{\var}[1]{{\normalfont\textit{#1}}}
\renewcommand{\paragraph}[1]{\bigskip\noindent{\textbf{#1}}\medskip}

\begin{document}
\pagenumbering{roman}
\setcounter{page}{1}
\thispagestyle{empty}
\begin{center}
\vspace*{-1in}
Argonne National Laboratory \\
9700 South Cass Avenue\\
Argonne, IL 60439

\vspace{.2in}
\rule{1.5in}{.01in}\\ [1ex]
ANL/MCS-TM-250 \\
\rule{1.5in}{.01in}

\vspace{1.5in}
{\large\bf NEOS Server 4.0 Administrative Guide\footnote{This 
work was supported by the Mathematical,
Information, and Computational Sciences Division subprogram of the
Office of Advanced Scientific Computing Research, U.S. Department of
Energy, under Contract W-31-109-Eng-38, and by the National Science
Foundation (Challenges in Computational Science) Grant CDA-9726385 and (Information Technology Research) Grant CCR-0082807.}
}

\vspace{.2in}
by \\ [3ex]

{\large\it Elizabeth D. Dolan\footnote{Electrical and Computer Engineering Department, Northwestern University, and
Mathematics and Computer Science Division, Argonne National Laboratory, Argonne, IL 60439; e-mail {\tt dolan@mcs.anl.gov}}
}

\thispagestyle{empty}

\vspace{1in}
Mathematics and Computer Science Division

\bigskip

Technical Memorandum No. 250

\vspace{1in}
May 2001
\end{center}

\newpage
\pagenumbering{roman}
\setcounter{page}{2}
Argonne National Laboratory, with facilities in the states of Illinois and Idaho, is 
owned by the United States Government and operated by The University of Chicago under the provisions of a contract with the Department of Energy.

\vspace{1in}

\begin{center}
DISCLAIMER
\end{center}

This report was prepared as an account of work sponsored by an agency of the United States Government. Neither the United States Government nor any agency thereof, nor The University of Chicago, nor any of their employees or officers, makes any warranty, express or implied, or assumes any legal liability or responsibility for the accuracy, completeness, or usefulness of any information, apparatus, product, or process disclosed, or represents that its use would not infringe privately owned rights. Reference herein to any specific commercial product, process, or service by trade name, trademark, manufacturer, or otherwise, does not necessarily constitute or imply its endorsement, recommendation, or favoring by the United States Government or any agency thereof. The views and opinions of document authors expressed herein do not necessarily state or reflect those of the United States Government or any agency thereof, Argonne National Laboratory, or The University of Chicago.

  \pagestyle{plain}
\newpage
  \tableofcontents
\newpage
\clearpage

\pagenumbering{arabic}
\setcounter{page}{1}
\pagestyle {plain}
\vspace*{1in}
\begin{center}
{\large\bf NEOS Server 4.0 Administrative Guide} \\ [2ex]
by \\ [2ex]
{Elizabeth D. Dolan} \\ [6ex]
\addcontentsline{toc}{section}{Abstract}
{\bf Abstract}
\end{center}
The NEOS Server 4.0 provides a general Internet-based
client/\linebreak[0]server as a link between users and
software applications.  The administrative guide covers
the fundamental principals behind the operation of the
NEOS Server, installation and trouble-shooting of the
Server software, and implementation details of potential
interest to a NEOS Server administrator.  The guide also
discusses making new software applications available through
the Server, including areas of concern to remote solver 
administrators such as maintaining security, providing usage 
instructions, and enforcing reasonable restrictions on jobs.
The administrative guide is intended both as an introduction
to the NEOS Server and as a reference for use when running 
the Server.

\newpage
\pagebreak\section{Introduction}
Consider hundreds of software applications with differing kinds of 
input data and
potentially thousands of users, each needing some of these
applications. The programming effort required for all software
developers to write application service providers that would make 
their codes remotely available over the Internet would be prohibitive.  
Consequently, the only way for users to access many applications is to
 download and install the codes locally. 
In some cases this is the best solution, but in other cases the code 
is either not
portable or is highly optimized for a given architecture. In these cases 
users might obtain an account on the host machine to use the software. 
Obviously, this solution is not scalable.

The solution that the NEOS Server provides is that of a general Internet-based
client/\linebreak[0]server, providing a link between users and software 
applications. Any application that can be tweaked to read its expected 
input from one or more files and 
to write meaningful results to a single file can be automatically integrated
into the NEOS Server. NEOS solvers are then available to users via all of 
the NEOS Server's interfaces.  The user interfaces provided by NEOS
are tailored to each particular solver based on information provided
by the solver administrators, yet the interfaces still possess the 
same look and feel
regardless of which solver the user is accessing. The advantages of
this system are
obvious in that users can freely browse the collection of solvers
without having to learn how to use a new interface for each software 
application.
Software developers benefit by easily adding their applications to 
a system that manages the complexities of client/\linebreak[0]server 
interactions for them.

In short, the NEOS Server handles interactions between a set of users
and a set of solvers.  A user submits formatted data to the NEOS Server
through one of its interfaces.  A \emph{solver\/} reads 
its input from one or more files, processes the input, and outputs the
results to a single file. For example, one solver may be a suite of
Fortran code that minimizes an objective function. In this case the
input could be a set of files containing the objective function and 
starting point source code, while the output could be a file containing the
solution vector and other information associated with the solution
process. The job of the NEOS Server is to connect the solver with
the user's data and the user with the solver's results.

The process just described is typically referred to as 
client/\linebreak[0]server.  Implementing client/\linebreak[0]server 
technology is straightforward, if somewhat tedious. Implementing a 
system abstract enough to handle the communications
needs of a wide range of applications and users while enabling as many 
conveniences as possible presents some challenges.  This guide
introduces the NEOS Server as a way to meet these 
challenges.  The guide begins with descriptions of the building 
blocks of Server functionality and the steps necessary to install 
and run the Server and solvers.  Descriptions become increasingly 
detailed as the guide progresses to the implementation of the 
NEOS Server in terms of submission flow and hints for 
administrative trouble-shooting and adaptation.

\subsection{Key Concepts}

The NEOS Server relies on the implementation of two main abstractions---that 
of data and that of services.  
The NEOS Server user interfaces currently consist of Email, World Wide Web, 
submission tools (TCP/\linebreak[0]IP sockets), and Kestrel (CORBA) 
interfaces.  To avoid needless complexity, each interface must provide 
the NEOS Server with the user's data in a consistent manner.  Indeed, 
the submission parser should have no need to distinguish among 
possible original user interfaces.  Further, the parser
should be able to accept various kinds of data to support a variety
of application needs.  To these ends, the NEOS Server has adopted a 
standard representation for all user data, described in Sections
\ref{tokens-general} and \ref{tokens-specific}. 

The Server's abstract implementation of services allows it to
schedule all user jobs as service requests, handled through a 
\script{solver} script.  At a central level, the Server need not 
be aware of the type of service it is enabling or 
whether user jobs are served on the Server machine itself or on a 
remote machine.  The NEOS Server assigns each job to a \script{solver}
script that executes application software via
requests to a communications daemon assigned to the application
and solver station.  The Server can delegate the tasks of interpreting
data and executing software to the (potentially remote) solver machines
through the NEOS Comms Tool, introduced in Section \ref{comms-intro} 
and detailed in Sections \ref{solver-script} and \ref{socket-requests}.
Through abstraction and delegation, the NEOS Server can enable
a wide variety of application services.

\subsection{The Token Configuration File}
\label{tokens-general}

Token configuration files are provided by solver administrators 
when they add their solver to the NEOS Server. 
Each line in a token configuration 
file specifies the name of an input file that the solver is 
expecting, along with tokens 
that can be used to delimit the data written to this file. 
A simple example configuration 
file  helps to illustrate:

\begin{verbatim}
Your 1st file::begin.a:end.a:A
Your 2nd file::begin.b:end.b:B
Your 3rd file::begin.c:end.c:C
\end{verbatim}

Notice that there is one line for each input file the solver is
expecting, namely files A, B, and C. In addition, there is a set of
tokens used to delimit the data coming from the user, and a label for
the data (e.g., ``Your 1st file''). Users of graphical interfaces
(e.g., submission tools, WWW) will simply see the label and a space for
entering the location of their local input file.  
Email users are expected to place all of their data into one file and 
delimit input sections with the appropriate tokens.  Cut-and-paste outlines for
email submission are available via email and through a Web page.  For example,
a token-delimited submission might look like the following:

\begin{verbatim}
begin.a
  contents of 1st file
end.a
begin.b
  contents of 2nd file
end.b
begin.c
  contents of 3rd file
end.c 
\end{verbatim}

The reason graphical interface users are relieved of token
delimiting their input themselves is that the interface 
does this for them before
sending the submission to the Server. Kestrel users have a client
that extracts their modeling environment's native format for NEOS 
submission, token-delimits the various types of data,
and sends it to the NEOS Server.  Kestrel users have no 
need to see the content between tokens, much less the tokens 
themselves.  In the end, though, the NEOS Server always receives data in 
token-delimited submission format, regardless of the interface that sent it.

When the Server receives the submission file containing the contents of the
user's three files, it must take this data and place each datum in the
files expected by the solver (i.e., \file{A}, \file{B}, \file{C}). 
The Server uses the solver's token configuration file to decide what 
to do with this input. In this case, all data between the tokens 
``begin.a'' and ``end.a'' gets saved to file \file{A}, and so on.  
The NEOS Server software packages the files, sends them to the machine 
housing the solver, and there unpacks the files.  It then calls the 
solver, which in turn reads in the files \file{A}, \file{B}, and \file{C}, 
processes the data, and generates a result. The result is then 
returned to the user via the same interface used originally to 
submit the data.

\subsection{The NEOS Comms Tool}
\label{comms-intro}

The heart of client/\linebreak[0]server technology lies in specifying
the types of communications available to the clients and servers.
While less obvious than the relationship between users (clients) and
NEOS (server), the dealings between the NEOS Server and the various
solvers also represent a client/\linebreak[0]server model.  
In the early days of NEOS, the Server and solvers were all on
the same system, so communication took place through files.  
Now, however, solvers are remote relative to the Server, for both 
security and modularity.  Hence, solvers must also have an Internet
interface to the Server, just as users do. This interface is 
referred to as the NEOS Comms Tool, which downloads the user files from 
the NEOS Server, invokes the solver application, and returns the 
solver's results to NEOS.

The NEOS Comms Tool is a small client/\linebreak[0]server application 
whose client is the NEOS Server and whose server is a daemon running on each
solver station. The NEOS Server, upon receipt of a job 
for a particular solver, connects to the machine running the Comms
Tool daemon for that solver, uploads the user's data,  and requests invocation of
the remote solver.  The Comms Tool daemon, upon receipt of this message, 
downloads the user's data and invokes the application software.  
Intermediate job results can be streamed back over the connection to the 
Server and from there, in turn, streamed back to the user. The Comms 
Tool daemon sends final results to the NEOS Server upon completion 
of the job, and the Server formats and forwards the results to the user.

NEOS users and solvers are thus completely abstracted
from one another via the NEOS Server. The advantages of this 
abstraction are immediate in that small changes in the solver 
applications leave the users unaffected.  On the other hand, 
improvements to the NEOS Server in the center of this
communication can benefit all solvers and users at once.  An example
of this advantage might be the addition of NEOS Server 
routines to handle data submitted to the Server with DOS or 
Macintosh file formatting.  Through a single change to the Server, 
users gain the ability to submit jobs from diverse systems to any 
of the solvers.  Solver users only need to know how to connect to 
the Server to submit jobs, and solver administrators only need 
to know how to make the Server aware of their solvers.

\pagebreak\section{Installing the Server}

The NEOS Server is intended to run on standard Unix-based platforms.  
Because almost all of the Server is written in Perl, 
the system have Perl version 5 or greater.  The NEOS Server relies 
on the \command{tar} and \command{gzip} facilities to package 
data for sending across the Internet.  To use the Email interface to 
NEOS, the system needs the \command{MH} mail facility 
and a \command{C} compiler installed.  

Before proceeding to the configuration and modification
of the Server, it helps to have answers to the following questions:

\begin{itemize}

\item{What userid do you want to run the Server under?  We suggest 
creating a new Unix account that will 
be dedicated to running the Server (e.g., the user \command{neos}). 
You can run the Server under your normal userid, 
but you will not be able to use the email facility, unless you want 
the Server treating all of your incoming 
mail as job submissions.  We do not recommended running the 
Server as root either.} 

\item{Where do you want the Server installed?}  
We recommend unpacking the Server somewhere on 
the local hard drive of the machine hosting it so
that the Server is not as prone to the difficulties of a network file system.
You must also choose a parent directory for the variable length files, 
including job submissions, logs, databases, and other records.
We refer to this directory as \var{server\_var}.  

\item{What interfaces do you want to support?  We suggest 
starting with the WWW interface.  Then you can see some 
of your progress installing the NEOS Server.}

\item{If you plan to support a WWW interface, what are the absolute directories and base URLs for your HTML 
and CGI pages?   You can get this information from your system administrators 
(e.g., absolute directories \file{/home/www/neos} and  \file{/home/www/cgi-bin/neos} with URLs \url{http://www.mcs.anl.gov/neos} and 
\url{http://www.mcs.anl.gov/cgi-bin/neos}).}

\end{itemize}

After deciding on the userid and home directory for the Server, 
become that user, go to that directory, copy the server package there,
and execute the following commands:

\begin{itemize}
\item{\command{gzip -d server-4.0.tar.gz}}
\item{\command{tar -xfv server-4.0.tar}}
\end{itemize}

You should have created the directory \file{server-4.0}.  For the remainder
of this manual, server file locations given are assumed to reside 
under \file{server-4.0}, unless they are listed as being in 
\var{server\_var} or some other directory supplied by the Server
administrator in the interactive Server configuration process. 
The next section will discuss configuration of the Server. 

\subsection{Configuring and Building the Server}
\label{configure}

To configure and build the Server, first make sure that you are 
logged in as the user that will
be running the Server. Then \command{cd} into \file{config/} and 
execute \command{make}.  
The \file{Makefile} will launch the configuration script 
\file{server/bin/}\script{config.pl} that will step you through 
the entire configuration process interactively.  
After all configuration parameters have been set, the script 
builds a set of front-end administration scripts that 
are used to start, stop, monitor, and maintain 
the Server.  A successful build looks something like this: 

\begin{verbatim}
Building Server:

building checker script...ok
building monitor script...ok
building restart script...ok
building kill-server script...ok
building cleaner script...ok
building report script...ok
building dailyreport script...ok
building weeklyreport script...ok
compiling some... 
make[1]: Entering directory `/neos/server-4.0/server/src'
cc address.c -o ../bin/address
make[1]: Leaving directory `/neos/server-4.0/server/src'
generating crontab for backup and reports...ok
\end{verbatim}

All of the configuration parameters are saved to the file
\file{server/lib/config-data.pl}.  The \script{config.pl}
script rewrites the contents of that file in the format
of a shell script and appends a call to the corresponding
Perl script under \file{server/bin/} for each administrative
script built.  The administrative scripts that can be executed
on the command line and used to run the Server 
are placed in \file{bin/}.  If these scripts fail to execute 
correctly, you may need to reconfigure the Server.

\subsection{Reconfiguring the Server}

If your Server is already running, be sure to first stop 
the Server by executing the administrative script 
\file{bin/\script{kill-server}}.
You can then reconfigure the Server simply by running \command{make} 
from within the directory \file{config/}.
You will notice that your old configuration has been saved and is 
now represented as the default.  
To remove your old configuration, run \command{make clean} before 
running \command{make}.  You can run 
\command{make} and \command{make clean} without losing any of the 
Server state, archived jobs, or logs saved 
in the directory \var{server\_var}.  You can remove these 
files before reconfiguration by executing
\command{make allclean}, which is equivalent to installing a fresh Server.
After reconfiguring your Server you can restart the Server with 
\script{bin/\linebreak[0]restart}. 

\subsection{Modifying Server Content}

Most text and html content used by the Server can be modified to 
suit the particular installation
of the Server. You may add or remove solver types in the 
file \file{\var{server\_var}/solver\_tree},
which  contains a list of solver categories (of which there 
must be at least 1).  Under \file{server/lib/html/} and 
\file{server/lib/txt/} you can modify the content of the 
Web site pages or the text information files associated with
the Server.
If you do modify any of the files, you will need to 
rebuild your Server to
reflect these changes in the automatically generated html 
pages and cgi scripts.  Rebuilding the 
Server can be accomplished by executing \command{make} in the 
directory \file{config} and accepting 
all of the defaults.  

\pagebreak\section{Running the Server}
\label{running}

If the Server has been configured properly, starting the
Server is as simple as typing a single command.  Keeping
the Server running and checking that it is executing as intended
requires a bit more effort.

\subsection{Starting and Stopping the Server}

To manually start the NEOS Server, execute the script 
\file{bin/\script{restart}} on the machine
where the Server has been configured.  You should see output similar to 
the following:  

\begin{verbatim}
killing receiver...
killing scheduler...
killing socket-server...
starting NEOS Server receiver...
starting NEOS Server scheduler...
starting NEOS Server socket-server...
\end{verbatim}

Assuming everything started correctly, the command \command{ps -o "pid args"} 
should generate output similar to the following:

\begin{verbatim}
5 /bin/perl /neos/server-4.0/server/bin/receiver.pl 
7 /bin/perl /neos/server-4.0/server/bin/scheduler.pl 
8 /bin/perl /neos/server-4.0/server/bin/socket-server.pl 3333 
\end{verbatim}

These are the three primary daemons that compose the Server.  The 
\script{receiver} script waits for incoming Email, WWW, submission tool,
and Kestrel jobs, passing these along to the \script{scheduler}.  The 
\script{scheduler} schedules these jobs on solver workstations by 
contacting the remote Comms Tool daemon running on the workstation.  
The \script{socket server} serves all TCP/IP socket traffic 
between the submission tool and the Server and the Comms Tool
and the Server. If any one of these three scripts is not running, the 
Server will not function correctly. 

To manually stop the Server, execute the script \file{bin/\script{kill-server}}.  
You should see the following output:

\begin{verbatim}
killing receiver...
killing scheduler...
killing socket-server...
\end{verbatim}

\subsection{Running the Server via Cron}

When you configure/\linebreak[0]build the NEOS Server, the file 
\file{config/\linebreak[0]crontabfile\linebreak[0]} is created.
This file contains crontab entries for the scripts \script{checker}, 
\script{cleaner}, \script{dailyreport}, and \script{weeklyreport}.  
The \script{checker} will restart your Server 
in the case of machine reboots or Server crashes.  
If added to the user crontab, usage reports created by the 
\script{dailyreport} and \script{weeklyreport} scripts are mailed
to the Server administrator and the email address for Server comments
at the appropriate time intervals.  The \file{crontabfile} also handles
calling the \script{cleaner} to archive old job directories, 
delete out-of-date Web results pages, and fill any empty areas
of the master database for currently running or strangely aborted jobs with
white space.  While the reports are optional, and restarting the Server is
only occasionally necessary, the Server administrator
should make a point of running the \script{cleaner} regularly.

If you want these scripts enabled via \command{cron}, 
you need to add the
cron entries in \file{config/crontabfile} to your current 
crontab.  If your crontab is empty, you can 
run \command{crontab crontabfile} to install the new crontab; otherwise,
 you will 
need to edit your crontab and manually add the new entries. On most systems, 
\command{crontab -e} opens an editor so that you can  edit 
your crontab directly.

Via the \script{checker}, the cron daemon will restart the Server after 
system reboots or in the event that the Server is not responding. 
The cron daemon invokes the checker, which connects to the 
\script{socket server} daemon and waits for a response.  If the 
\script{socket server} daemon is not running (or
does not respond in time), then the \script{checker} will assume that the 
entire NEOS Server is dead, in which case it will kill the remaining 
Server daemons and exit. Upon the next invocation of the \script{checker}, the 
Server will be restarted via \command{bin/restart}.

The cron daemon does start processes as the
user identity that invoked the \command{crontab}, but cron jobs
entail minimal initial running environments.  Section \ref{solver-driver}
contains more on the limitations of cron for solver 
administrators.  The Server administrator should know that 
if the Server is restarted by the cron daemon, the \command{ps} 
command will no longer show the running daemons (as they no longer have a 
controlling terminal).  Check the manual pages on your system
to find the option to \command{ps} that allows you to view processes
without controlling terminals.  You should also run a few other tests
to check your Server.

\subsection{Server Checks}

Once you have determined that the daemons are all running,
several other tasks will help to show that everything is 
working correctly.  A simple job submission from any interface
can test most of the main Server scripts.  Submissions via Email
and from the Web site interface, if enabled, provide assurance
that the special features associated with these interfaces are
operable.  Section \ref{monitoring} begins the discussion of 
additional tools to help debug any problems.

In some ways the submission tool interfaces represent the best 
initial test of the Server because a successful submission 
tool job completion indicates that socket requests are flowing
through the \script{socket server} daemon and the \script{receiver}, 
\script{initializer},
\script{parser}, \script{scheduler}, and \script{solver} scripts 
are at least minimally functional.  First you must install one of the 
submission tools (Tcl/TK or Java, currently).  More information
on installing the Tcl/TK client tools is given in Section \ref{start-comms}.
  If you
are able to view the tool's graphical user interface, then the 
\script{socket server} script is generally functional. 
You can submit a blank \emph{Adding/\linebreak[0]Modifying a Solver} 
job from the submission tool to test the functionality of the
job requests to the \script{socket server}.  The solver should return
a list of the categories in \file{config/solver\_tree}.  

You can submit a simple Email interface job
to the Server by mailing the following message to the appropriate 
address:
\begin{verbatim}
help admin:addsolver
END-SERVER-INPUT
\end{verbatim}
Requests for Email interface help are scheduled like 
other jobs, so results from this
submission should indicate the functionality of the Email interface
to the NEOS Server.  Additionally, the help returned includes a 
template for mailing \emph{Adding/\linebreak[0]Modifying a Solver}
jobs.

If you have opted to enable the Web interface, 
check the URL you gave to ensure that the Web site is visible.  If
you find nothing at the URL you expected, you should check the 
directory where you specified the Web pages should be built.  
If the pages are there, check with your Web master to determine the
correct URL and reconfigure the Server.  If you find no Web pages, 
double check your configuration
entry and check that you have write permission to the parent directory.
Assuming the Web site has been built, look for all of the solver categories
you added to the \file{config/solver\_tree} on the \file{server-solvers.html}
Web page.  Under \emph{Adding/\linebreak[0]Modifying a Solver}, go to
the \command{WWW Form} and resubmit the blank job to test most features
of the Web site CGI scripts.  If the other interfaces are working while
the Web interface is not, your Web master may be able to help you by giving
you any system-specific pointers about the use of CGI scripts.

Should any of the above tests fail, you may find more specific hints
as to the problem by monitoring the Server logs and database.  The
socket log can be especially helpful when the Server fails to restart
properly.

\subsection{Monitoring Server Traffic}
\label{monitoring}

\begin{figure}[hbt]
\centerline {\psfig{file=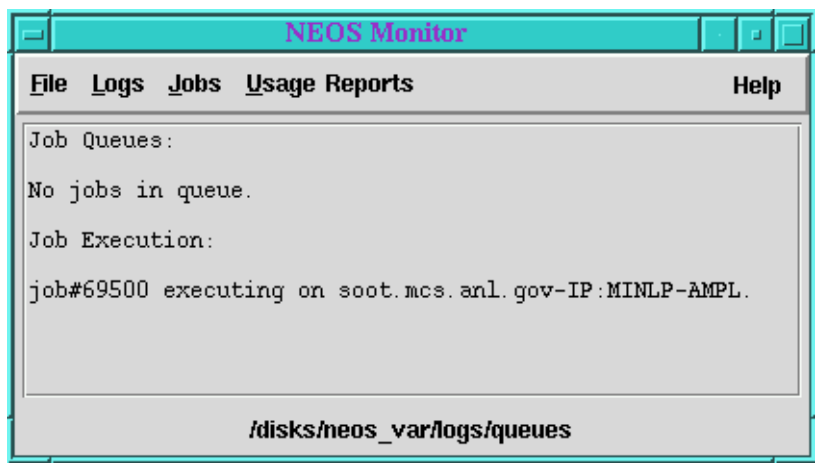, height=2.375in}}
\caption{NEOS Monitor}
\label{monitor-gui} 
\end{figure}

While all of the Server logs and databases are essentially just
files, we do provide an extra \command{monitor} tool with which
you may view and query them.  The monitor is not a requirement for
running the Server, and the system must have \command{wish} and
\command{Tcl/TK} installed for the monitor to function.  To start
the monitor, run the \file{bin/\script{monitor}} script.  The script
should open a graphical user interface (Figure \ref{monitor-gui})
with menu options to view the various Server logs, make queries to 
the master database, retrieve job submissions and results, and
create usage reports.

All Server activity is logged in the directory \file{\var{server\_var}/logs}. 
You should see one file for the \script{receiver} daemon, one 
for the \script{scheduler}, one 
for the \script{socket server}, and one for the \script{checker}.  
Additionally, there is the file \file{queues},
which is not really a log, but lists the currently executing jobs 
and jobs in the scheduler's execution
queues.  

Assuming that everything was configured correctly and the Server 
started, the receiver log should contain output similar to:
\begin{verbatim}
receiver: My Server restart on Aug 8 16:38 PDT 1999.
\end{verbatim}
The scheduler log should contain output similar to
\begin{verbatim}
scheduler: Restart on Aug 8 16:38 PDT 1999.
\end{verbatim}
The socket log should contain output similar to
\begin{verbatim}
My Server accepting connections on opt.mcs.anl.gov port 3333.
\end{verbatim}
The checker log (if the Server was started via cron) should contain 
output similar to
\begin{verbatim}
client.pl: ERROR: connect: Connection refused
checker: killing My Server processes on Aug 8 16:38 PDT 1999.
checker: restarting My Server on Aug 8 16:38 PDT 1999.
\end{verbatim}
and the queues log should look like this:
\medskip

\noindent
\command{Job Queues:}
\smallskip

\noindent
\command{No jobs in queue.}
\smallskip

\noindent
\command{Job Execution:}

\smallskip

\noindent
\command{No jobs executing.}
\bigskip

A couple of hints about these logs are worth a Server administrator's
attention.
If you ever have the need to completely abandon jobs listed in the
\file{queues} (because of some catastrophic failure), 
kill the Server, delete \file{queues}, and delete
every file under your \file{\var{server\_var}/proc/\linebreak[0]stations/} directory
before restarting the server.  If you want to reduce the logs' sizes
by deleting old entries or deleting the logs entirely, be sure to
restart the Server afterward.  Most of the NEOS Server
daemons are unable to write to altered logs, eliminating a good 
source of trouble-shooting data.

One common problem that can cause Server tests to fail is
easily spotted in the socket log.  When the NEOS Server is not
killed correctly, you may sometimes see the restart message
\begin{verbatim}
Cannot bind to port 3333!
\end{verbatim}
After multiple attempts to bind the port, the NEOS Server will send
the Server administrator an email message warning that the Server
could not be started.  If you make changes to the Server configuration
that do not appear to be functioning after you restart the Server,
you should check the socket log to see whether an older version
of the \script{socket server} might still be running.  In fact, it
is not a bad idea to check the socket log every time you restart the
Server.

\pagebreak\section{Solvers on the NEOS Server}
\label{solvers}

One of the conceptual challenges presented by the
the creation of a generic application service provider involves
the interface the NEOS Server supplies for communicating with 
a wide variety of software applications.  The interface should
be flexible enough to handle many different input, output, and
systems demands on the part of the applications while adding 
enough support to make the process of adding an application to 
the available pool worthwhile.  Our system is designed to 
allow solver administrators at distributed locations to take
responsibility for solver applications that run on
their sites.  The solver administrators,
like other users, need not have accounts on the same system
as the NEOS Server.  They register their solver with the 
NEOS Server using an administrative solver, write an application 
wrapper that can be
executed on their site, and start a communications daemon
on each solver station they make available.  The NEOS Server
is designed with few requirements to make an application
accessible through the system but with many options for customizing
and enhancing solvers.

Section \ref{solvers} offers solver administrators an 
explanation of the steps involved in adding applications to
the NEOS Server.  
The NEOS Server includes a collection of built-in 
administrative solvers, which are used in registering
new solvers.  The information required for registration
is detailed along with supplementary data helpful to
solver administrators for informing users about the 
purposes and requirements of their
software.  Particular attention is paid to the role solver 
registration plays in building the NEOS Web site.
The necessary execution requirements for software applications
are presented in Section \ref{solver-driver} on hooking 
solvers to NEOS. Finally, Comms Tool daemon startup on the 
solver stations completes the picture of building a NEOS Server 
by adding solvers.  Along with these sections of the guide, 
the NEOS Server administrator can 
point solver administrators toward the \emph{Solver Administrator
FAQ}, \url{admin_faq.html}, that is built with the NEOS Web site.

\subsection{Administrative Solvers}
\label{admin-solvers}

The NEOS Server comes packaged with a suite of administrative solvers,
which function in much the same way as other solvers. 
These solvers include \emph{Adding/\linebreak[0]Modifying a Solver\/} 
\mbox{(\solver{admin:addsolver}),} 
 \emph{Enabling/\linebreak[0]Disabling a Solver\/} \mbox{(\solver{admin:status}),}
\emph{Kill Job\/} \mbox{(\solver{admin:}}\linebreak[4]\mbox{\solver{kill\_job}),} and the 
\emph{NEOS Help Facility\/} \mbox{(\solver{admin:help}).}  The first two 
solvers are used strictly by solver administrators while the third 
may also be employed by a user with the appropriate job password.  
The help facility provides Email interface information to a user about a 
particular solver.

The most important two of these administrative solvers provide the
mechanisms for adding and deleting solvers.  The 
\solver{admin:addsolver} solver allows NEOS to 
automatically increase its collection of solvers. The 
basic purpose of the \solver{admin:addsolver} is to generate 
all of the user interfaces to solvers based on the information
provided by the solver administrators. For example, when a 
solver is added, \solver{admin:addsolver} will enter the 
solver identifier in the NEOS Server's list of available 
resources and automatically create the solver's WWW 
interfaces (both HTML and back-end CGI). The \solver{admin:status}
solver allows the Server to shrink its solver collection.
If solver administrators ever wish to temporarily disable their solver 
or completely delete their solver from the Server, they can just 
submit this job request to the \solver{admin:status} solver, 
which deletes the solver identifier from the list of available 
resources and, using the WWW example, either temporarily disables the 
links to the solver's Web pages or completely deletes the Web 
area for the solver.  

Solver administrators can read more about the other administrative
solvers in Section \ref{admin-imp}.  That section on the implementation
of administrative solvers describes how \solver{admin:help} takes
advantage of \file{email-help} information registered with the
solver.  \emph{Kill Job} should be tested by solver administrators
before they decide whether to enable it for their solvers, as
discussed in Section \ref{start-comms}.  Neither of these solvers
is crucial to the NEOS Server, but they do offer features
helpful when the Email interface is enabled or when
remote solver stations are able to completely kill off application
processes on demand.

The \solver{admin} solvers available to 
solver administrators allow them the greatest 
possible flexibility in determining the interactions between 
the NEOS Server and their solver without requiring direct
access to the machine hosting the NEOS Server.  The NEOS Server
makes its administrative solvers available through all of the 
interfaces provided for regular solvers so that solver 
administrators have no need for an account on the NEOS 
Server machine.  These interfaces are built based on the same
type of token configuration files required for other solvers.
The main difference between the administrative solvers and
other solvers is simply that the administrative solvers come
packaged with the NEOS Server.

\subsection{Registering a Solver}
\label{registering}

Solver administrators can register their solvers with the NEOS Server
through any interface by using the administrative solver 
\emph{Adding/\linebreak[0]Modifying a Solver\/} 
(\solver{admin:addsolver}).  When a solver is registered, it is 
automatically enabled, meaning the NEOS Server may attempt to 
relay incoming submissions to the solver.
Through the \emph{Enabling/\linebreak[0]Disabling a Solver\/} \mbox{(\solver{admin:}}\solver{status})
solver, administrators may temporarily disallow submissions to their
solvers and enable the solver again later.  The registration form
submitted to \solver{admin:addsolver} contains entries for 
the following information.

\begin{itemize}
\item \textbf{Solver Type}

New solvers must be placed under an existing NEOS Server solver
category.  Submitting a job to \solver{admin:addsolver} with 
a blank or incorrect solver type will trigger the Server to 
return a list of available solver categories.  The 
\solver{admin:addsolver} expects to receive the abbreviated
type name (e.g., \command{misc}, not \command{Miscellaneous}).

\item \textbf{Solver Identifier}

The NEOS Server identifies solvers through their type,
identifier, and password. A solver identifier should be chosen as a 
single word without any special characters, guaranteeing only that
letters, numbers, and underscores will function correctly.  
A solver identifier must be unique within its solver type category.

\item \textbf{Solver Name}

Each solver is also given a full name, which may contain
spaces and other special characters.  The NEOS Server generally
displays the solver name in its various interfaces but uses the
solver identifier internally.  

\item \textbf{Solver Password}

The password entry represents an attempt to ensure that only 
authorized users or the Server administrator is able to reconfigure 
solvers.  This password is not extremely secure. Solver administrators 
should \textbf{not} use their regular account passwords.

\item \textbf{Contact Person}

The \solver{admin:addsolver} scripts insist that solver administrators
give an email address so that they can receive error messages 
when their solver misbehaves or is abused.  The address is not currently
checked for validity but may be in future releases of the NEOS Server
package.

\item \textbf{Workstations [jobs allowed]}

The solver administrators must also give a list of workstations where their
solver will run.  The NEOS Server must have the full machine address of 
each solver station.  Optionally,
the administrator may enter a positive integer on the same line following 
the machine address that indicates the number of jobs submissions that can
be run on the solver station concurrently.  An entry might look like the
following:

\command{harkonnen.mcs.anl.gov 3}

\item \textbf{Token Configuration}
\label{tokens-specific}

The solver administrator gives the full path to a file containing
the tokens that will delimit job submission data for the solver.
Each line in the configuration file contains five colon-separated entry fields.
The first field provides the data label that appears on the Web and submission 
tool forms for the solver.  The second field specifies the type of 
data or the default value of the data.  The data types that
may be specified are 
\command{TEXT}, \command{BINARY-ON}, \command{BINARY-OFF}, and 
\command{RADIO}.  The \command{TEXT} type referred to here provides a large
text entry field with scroll bars in the graphical interfaces.
Any other entry will be treated as a default value 
for a text variable or file name and appears only in the
submission tool GUI. The third and fourth fields specify beginning and 
end tokens.  If the end token is \command{NULL}, the data type will 
be a simple text variable (without scroll bars in the GUIs), and the
submission format will require that the value be given on the same
line as the beginning token.  If the beginning token for a file
name data type (i.e., an unspecified type with an end token)
ends with \command{BINARY}, the file uploaded 
to the NEOS Server will be preserved with no mangling.  The fifth
field gives the name of the file where the data will be written
on the solver station.  Lines in the token configuration file
appear as follows.
\begin{verbatim}
Objective source: fcn.c: begin.func: end.func: FCN
Language: RADIO .C,c .Fortran,fortran: lang: null: LANG
Number of variables: : NumVars: NULL: NumVars
Use alternative algorithm?: BINARY-OFF: AlgB: null: AlgB
Job label: TEXT: BEGIN.COMMENT: END.COMMENT: COMMENTS
\end{verbatim}

\hspace{12pt} The token configuration file is the only
file necessary to register a solver.  It allows various types
of data to be written to specified files.  In this example, 
the Server parses incoming submissions to the solver 
and writes the source code for the function evaluation 
to \file{FCN}.  It places the text value \command{c} or \command{fortran}
in the \file{LANG} file and another integer text value
in \file{NumVars}.  The \command{BINARY} data type
accepts values of \command{yes} or \command{no} and writes
them to \file{AlgB}.  Finally, a text field for comments saves
its data in the file \file{COMMENTS}.  The begin and end tokens
are not case sensitive, but files appear by exactly 
the fifth field names in the directory where the solver driver or executable
is invoked.  A fully token-delimited submission to a solver with
identifier \command{mine} and type \command{misc} registered with
this token configuration,
whether formatted by the interface or directly by the user as
an email submission, might appear as follows.

\begin{verbatim}
TYPE MISC
SOLVER MINE

begin.func
double fcn(double *x, int n) {
    double f = 0.0;
    for(int i=0; i<n; i++)
        f += x[i] * x[i];
    return f;
}
end.func

lang = c
NumVars  = 10
AlgB     = no

begin.comment
A simple example.
end.comment

END-SERVER-INPUT
\end{verbatim}

The NEOS Server makes no effort to check the contents of the
input files for validity.  This task is delegated to the 
individual solver drivers (Section \ref{solver-driver}).  The files and other
data described next are optional, and further descriptions
of the Web-related entries follow in Section \ref{WebSite}.

\item \textbf{Usage Restrictions}

Solver administrators may limit the number of jobs sent to
their solver stations.
The usage restriction file contains lines of the form

\command{<users> <minute|hour|day|month> <limit>}.  

For example:
\begin{verbatim}
#max                hour  20 
#max_any_one_user   day   30
#max_any_one_domain month 100
\end{verbatim}

\item \textbf{Email Help}

The Email help file is sent to users who request help via email from
the \solver{admin:help} facility.  It should contain a template
with the data tokens for Email interface submissions.

\item \textbf{Submission Tool Help}

The submission tool help file is returned to users who
submit a request for help from a solver's submission form
as accessed through a submission tool.

\item \textbf{WWW Help}

The Web help file contains the text that appears on the solver's
Web submission page.  The text may contain HTML tags for
better formatting.  In addition to the usual HTML tags,
this file should contain one \command{<NEXT\_ENTRY>} tag
for each entry in the token configuration file.  A 
description of the data necessary for each entry may
precede the \command{<NEXT\_ENTRY>} token.  If the
solver administrator supplies no WWW help file, the
Web submission form is still created, with only the
labels from the first field of each token configuration
line to guide the user. 

\item \textbf{WWW Samples}

The Web samples file gives descriptions and URLs for sample solver 
submissions.  The submissions must be formatted as would
an Email submission.  The WWW samples file contains two
lines for every sample.  The first gives a description that
may contain HTML tags.  The second is the full URL to a 
Web-accessible sample.  The sample page under the solver's
home page on the Web site will enable users to make practice
submissions.  To provide easy viewing of the sample, a
hyperlink may be added to the description.  

As an alternative to the two-line format, which results in
a list of samples, the samples may be formatted in a table.
For example,

\begin{verbatim}
<COLUMNS 3>
Problem
Algorithm A
Algorithm B
<a href=''ftp://ftp.some.site/exmpA.txt''>Production</a>
http://www.some.where/exmpA.txt
http://www.some.where/exmpB.txt
\end{verbatim}

would create a table with examples for algorithms \command{A} and
\command{B}.

\item \textbf{WWW Abstract}

The WWW abstract file is displayed as the solver's home page
within the NEOS Web site.  The text may contain HTML
tags as would appear below the \command{<BODY>} token.
Knowledge of HTML is not necessary, but plain text may
not be formatted as nicely.

\item \textbf{WWW Background URL}

The Web background URL text field should contain the full URL to a Web page 
providing more information about the solver if such a page
exists.

\end{itemize}

\subsection{Taking Advantage of the NEOS Web Site}
\label{WebSite}

When registering a solver, the NEOS Server offers the opportunity
to craft a mini-Web site devoted to that solver.  
The NEOS Server Web site construction allows for a 
separate group of pages for each solver registered on the system.  
The mechanisms that allow automatic solver Web site construction
and enable Web interface are discussed in more detail in 
Section~\ref{WebServer}.

Each 
solver has a descriptive homepage, a sample submission
page, a user comments page, a page of instructions for email use with a 
solver-specific submission template for 
emailed job requests, and a form for submitting jobs from a Web browser.
The NEOS site links these pages via a hyperlink to the solver homepage 
from a central NEOS Web page listing all solvers by category.

When solver developers register their solver on the NEOS Server, they must
supply at least a solver type, identifier, name, contact email address, and a 
token configuration file.  From
this information alone, the NEOS Server can build the solver's Web submission
form and the pages describing the email template.  The solver's name will
appear on the list of available solvers with other solvers of that type.
In a technical sense, these tools are enough.  Unless all of the Server's
intended users are already familiar with the solvers offered, though, more
information is critical.

\paragraph{Sample Submissions}

\noindent
The quickest introduction a user will find
to submitting NEOS jobs is through the sample Web submissions.  
Solver administrators should pick a few appropriate trial submissions,
prepare them in token-delimited format for the NEOS Server, 
and place them in a Web-accessible area.  The extensions on the files 
should indicate plain text so that the Web browser will not 
try to render them as if they were HTML when users view them in a browser.
The administrators compile a file outlining the sample submissions 
available with the simple format of one line for the example's label 
followed by a line containing only the URL of the sample file.
The labels may include hyperlinks to the sample text.  
By providing
the user with working code, the sample submission Web page makes the use
of token-delimited entries clear.  It also allows the skeptical to prove
to themselves that the NEOS Server is up and running without investing
much time in learning the system.

\paragraph{Solver Homepage}

\noindent
The registration form asks for an optional WWW abstract.  This plain text
description of the solver and its usage on the Server is incorporated
into an HTML body to create the solver's homepage on the NEOS site.  
The process of
creating solver homepages around the developers' supplied abstracts allows
a similar look and navigational feel throughout the solvers offered.  The
name of the solver becomes a link to off-site Web pages when solver
administrators supply a background URL.
A \emph{Sample Submissions}
link steers users to the examples for each solver, and 
button links for sending in \command{Comments and Questions} or for
returning to the NEOS Server \command{Home} are woven into each solver homepage
as well.  

For
a more sophisticated homepage, HTML mark-up tags may be included in the developer's abstract text as they would
in the \command{<BODY>} section of a normal HTML file.  Because Web browsers
will interpret the abstract as part of an HTML page, certain special
characters must be delimited as for HTML.  For example, the \command{<} 
symbol is denoted
\command{\&lt;}, the \command{>} symbol is \command{\&gt;}, and the ampersand
symbol is \command{\&amp;}.  Often in practice the abstracts are merely 
pasted from the solver developer's descriptive Web page, requiring no changes.
Again, the solver administrators may devote only as much effort
as they desire to registering the solver.  Even if they submit
no abstract whatsoever, the NEOS Server still creates a minimal solver
 homepage with the appropriate links.

\paragraph{Web Interface}

\noindent
The same philosophy is true of the Web submission interface.  If 
solver administrators choose not to submit a Web help form for the
solver during registration, the Web form for job submission will 
be built based on the token configuration file.  If administrators
want to offer more detailed explanations or examples for the 
form fields, they need only create a text file with a \command{<NEXT\_ENTRY>}
token delimiting the desired location within the text of each entry in the 
token configuration file.  The resources of HTML hyperlinks are available
to administrators, but they have the option of placing their solver
on the NEOS Server with little effort just to see whether their software
will be useful to a larger community.  Administrators can always
reregister their solver, adding details over time.

\subsection{Hooking Solvers}
\label{solver-driver}

To run an existing application via the NEOS Server, the application
must be able to extract its input from a predefined group of files
and send its output to a file called
\file{job.results}, also in the current working directory.  Generally,
some sort of driver script is required to handle these tasks.  
Any executable script
that can start the application software so that it receives the user input 
and writes output to \file{job.results} in the current directory is 
acceptable.  A few points warrant consideration when writing 
such a script.  

When the NEOS Server sends a job to a Comms Tool daemon, the
daemon creates a temporary working directory for the job. 
The user-supplied files are unzipped in that directory and will be
named as specified by the fifth field of each token
configuration entry.  These user files will be local to the 
driver script when it is invoked, but the driver script should not rely on
relative locations for any other files.  The driver script should state 
absolute paths to the application software and supporting files.  
Also, full path information for environment variables
like the \command{PATH} should be set if the Comms Tool daemon is started
with the \command{Cron} option, which has an extremely limited environment
and does not check for any shell run command files (e.g., \file{.cshrc}).  
If the administrators' solver driver works when they start it but 
begins to fail later, these issues should be checked first in 
debugging.

Because applications may fail, it is in the solver administrators'
interest to ensure that the user receives as much information 
as possible during job execution.  Because users have intermediate 
results streamed into their Web browser or submission tool in 
approximate real time, the intermediate output from the application 
should be written to the Unix standard output without any buffering.  
The driver may need to flush output buffers explicitly to 
accomplish real-time updates and avoid the impression that the
application is stalled.  The final results should also be 
flushed as they are written to \file{job.results} so that failed 
applications provide the best possible hints about problems encountered.
A NEOS solver that offers a user comment field should write the comments 
to \file{job.results} before the driver invokes the application.  
Users who write comments to label their submissions can then 
identify a job in case the application software fails, is killed, 
or times out.  Execution information written to 
\file{job.results} can be overwritten with summary 
data by the driver after the job has completed successfully.  Simply 
renaming the application's default output file at the end of execution, 
though, may mean that no \file{job.results} file is created for 
terminated runs, failing to meet the Server's one requirement of 
solver drivers.  The user is more likely to resubmit jobs verbatim 
for which they receive no results than jobs that return error messages,
and administrators will receive warning messages from the NEOS Server
when their solver fails to create \file{job.results}.  

While output to \file{job.results} is technically all 
that is required of a solver driver by the NEOS Server, 
the sophistication of the driver script greatly determines the overall 
functionality of the application.
The manner in which an application reads options affects the 
amount of control users have over the application's execution.  
One common task of the solver driver is to specify command line 
options for the application, depending on the user input.  An 
application that reads in an options file may require less 
effort on the part of the driver than an
application that accepts only command line options.  
Depending on the options available, though, it may be a good idea for the
driver to scan the user options file before the application 
is executed regardless of whether the application can read the options
file itself.  In this way, solver administrators can check that 
no unreasonable options are being leveraged by the user.

The solver driver script also serves to limit
functionality not explicitly handled by the NEOS Server or via an option
in the solver registration or Comms Tool startup.  Not only 
should reasonable options be enforced, but user code that will be 
executed on a solver station should be scanned for potentially 
harmful commands.  Running the Comms Tool via the account of a 
specially limited user can eliminate many of the potential risks.  
Alternatively, administrators may be able to exploit special options 
maintained by some applications defining restricted use by, 
for example, disallowing shell calls or file I/O.  
When solvers are distributed across different systems with varying 
policies, often the only way to gain any access to a site is by 
imposing specific limitations on that access.  The next section
discusses the technical mechanism by which the NEOS Server accesses and
coordinates with other user identities on remote solver stations,
including some options to limit the application execution.

\subsection{Starting Communications with the NEOS Server}
\label{start-comms}

The NEOS Server supports a Perl implementation of a Comms Tool 
using TCP/IP for Unix.  More information about the specific requests that
can be made between the Server and Comms Tool daemon can be found in Section
\ref{socket-requests}.  While the Server specifically supports only one
implementation of a Comms Tool, others could be written to communicate 
with the NEOS Server from other platforms if they handled all of the requests
mentioned in Section \ref{comms-requests} and correctly made use of the
requests available to them as described in Section \ref{neos-requests}.
The information given next pertains to the Comms Tool implementation
supported the Server and its Tcl/\linebreak[0]TK graphical user interface
for Unix systems.

To take advantage of the NEOS Server-supported Comms Tool, solver
administrators must first install the Unix client on their 
site.  They should download the current \var{client-version} package 
for their system and unpack it in the location of their choice.  The
client should create a \file{client-\var{version}/} directory with a
\file{Makefile} in it.  Solver administrators should check the \file{Makefile}
to ensure that the machine name and port for the NEOS Server are 
correct and then \command{make} the client.  This installation process
should create \script{submit} and \script{comms} scripts under the
directory \file{client-\var{version}/bin/}.  The \script{submit} script
launches the Tcl/TK submission tool, and the \script{comms} script
launches the Tcl/TK Comms Tool interface.  If the Comms Tool graphical
interface does not load, then either the version of the client
in incompatible with the solver administrators' system and another should be
tried, or the client could not establish a socket connection with the
NEOS Server.  Error messages should help to determine which scenerio 
is the case.

The solver administrator must start a Comms Tool daemon on each of the
solver stations listed in the registration process.  Communications
cannot begin unless the solver is properly registered on the Server.
Because the NEOS Server communicates directly with the Comms Tool
daemons via a socket connection, no passwords are required.  However,
systems that rely on substantial security, such as those behind a 
firewall, may not allow normal users to bind to a socket for communication.
Also, while the Tcl/\linebreak[0]TK Comms Tool interface expects 
that administrators
can start Comms Tool daemons on all of their solver stations from one 
machine, in practice administrators may need to log on to each solver
station separately, giving a password, to begin communications.  The
Tcl/\linebreak[0]TK Comms Tool can interface across machines using 
\command{rsh} or \command{ssh}; more information on the use of
\command{ssh} and other aspects of the client tool is available
in the \emph{Solver Administrator FAQ}.

When solver administrators first execute the \script{comms}
script of an installed \script{client-\var{version}} tool, the
script creates a directory named \file{.comms} under the 
administrators' home directory on the system.  Alternatively,
the Makefile for the \script{client-\var{version}} tool may
be altered to set an environment variable, \texttt{CACHE\_HOME},
which will be the parent directory of \file{.comms}. 
Logs of communications
between the Comms Tool daemons and the NEOS Server \script{socket server}
are stored under the \file{.comms} directory, and the 
temporary working directories
for each job are created under this directory.  The process identifiers
for running daemons and the port associated with each are listed, and
backups of the 
commands that start the Comms Tool daemons are also stored here.

Information required by the Comms Tool to start a daemon 
includes the full name of
the machine hosting the NEOS Server and the port on which the 
\script{socket server} is listening.  Solver administrators must
be able to identify their solver by type, identifier, and password
and have access to the workstations that will serve as solver stations.
An email address is required, as is the absolute path to the 
solver driver script discussed in Section
\ref{solver-driver}.

The Comms Tool offers numerous options to customize the execution
behavior of solver applications.
Solver administrators may choose
to change the default time limit, which will work effectively only if
their operating system supports full use of process groups and the 
application software does not make any changes to the process group id of
a job.  The same restrictions apply to whether a solver's jobs may be
labeled \command{Killable} when starting communications.
The file size limit field may be adjusted to limit the size of
any file created by the application, including all intermediate and
results files, on systems that support the \command{limit} command.
When selected, the \command{Notify} option requests that mail
be sent to the contact address provided each time a job submission 
arrives.  The \command{Debug} option ensures that any messages sent 
to the standard error stream of the solver are piped back to the user.
\command{Save} prevents the Comms Tool daemon from deleting working directories
when a job is completed.  The \command{Cron} selection attempts to start
a cron job to check occasionally that the Comms Tool daemon for 
the solver is running on the solver station and restart it if necessary.
If solver administrators wish to disable this feature after communications
are started, the \command{Disable Crontab} button does so.  Solver
administrators should normally disable the crontab
for their solver before killing the Comms Tool daemon.  Otherwise, the cron daemon
will continually attempt to restart communications with the NEOS Server.

\pagebreak\section{Server Implementation}
\label{implement}

The remainder of this document is not intended for
the casual reader interested only in the goals and
use of the NEOS Server.  Rather, we describe
the software behind the Server for NEOS 
administrators who encounter problems, have special needs, 
or would like to prepare themselves for any eventuality.

\subsection{The Directory Structure}

The NEOS Server 4.0 package, when unpacked, places all
of its contents in a directory called \file{server-4.0/}.
The directory \file{server-4.0/} contains three subdirectories:
\file{bin/}, \file{server/}, and \file{config/}.  When the
Server is installed, it must also have access to some 
directories specified by the user, including \file{server\_var},
a directory for Web pages, and a directory for Web CGI scripts,
which may reside on separately mounted drives.  Most of our
discussion of NEOS Server implementation focuses on files
found in the subdirectories of \file{server-4.0/}.

The directory \file{bin/} contains front-end 
administrative shell scripts 
that make calls to some collection of Perl scripts within 
\file{server/bin/} after the shell 
environment has been set appropriately.  If you glance at 
the contents of any of the scripts
in \file{bin/}, you will notice they are all close to 
identical with the exception of the 
last line that executes one of the Perl scripts in \file{server/bin/}.  
These administrative scripts are all built during the configuration
of the Server; hence, \file{bin/} is empty when the Server is
first unpacked.

The directory \file{config/} contains the \file{Makefile} for 
installing the Server.  The \file{crontabfile} containing
entries for use by the cron daemon is built within \file{config/}
as part of the installation process.

Directory \file{server/} is
the only one at its level that is also a parent directory.
Its subdirectory \file{server/src/} contains one \command{C} 
source file, \file{address.c}, along with a \file{Makefile}
that is executed during Server configuration if the Email
interface is enabled. 
The directory \file{server/bin/} contains Perl scripts and 
other executables called directly by the Server after its 
environment variable values have been established during
configuration.  
The most extensive subdirectory of \file{server/}, \file{server/lib/},
acts as parent to most of the files that define the content, rather
than behavior, of the Server.

The \file{server/lib/} directory contains files generally considered 
read-only by the Server.  This directory does contain templates for
executable files, but these are not executables directly related to 
the Server, as is the case with all executables in  
\file{server/bin/}.  For example, the scripts under \file{server/lib/Perl/}
are downloaded to the remote solver stations and called by
the Tcl/TK Comms Tool daemon provided with the Server.  The scripts
under \file{server/lib/TclTK/} are downloaded by
the Tcl/TK clients, which are packaged separately but supported by the
Server, and executed on users' local machines to build 
the graphical interfaces for the submission tool or Comms Tool.  
Similarly, the Server keeps
an archive of Java classes for the Java submission tool under
\file{server/lib/java/} so that the \script{socket server} can send
them to the local machines of Java submission tool users in need of
updated client versions.   The exceptions to the read-only 
rule lie under \file{server/lib/admin\_solvers}, where files 
ending in the suffix \file{.server} are rewritten with Server-specific
data to a file in the same directory, named without the suffix.
If the Server administrator wants to change the content of some
aspect of the Server for which no option has been offered during
configuration, \file{server/lib/} represents a good place to look 
for the appropriate files.

The user-supplied directory \var{server\_var} contains files and 
directories created, expanded, and deleted in the course of Server
operation.  
This space includes directories for 
\file{databases}, \file{logs}, \file{jobs}, and
process ids of running processes (\file{proc}), a \file{spool} 
directory for incoming 
jobs (e.g., from the Web server), a \file{lib} directory containing 
mostly individual solver registration information, 
and a \file{tmp} directory for use by the 
various Server processes.  

\begin{table}[htb]
\caption{\textbf{Administrative Scripts}}
\label{admin-scripts}
\smallskip

\begin{tabular}{|l|l|}
\hline
Script & Purpose \\
\hline
checker.pl & invoked by bin/checker to ping server \\
cleaner.pl & invoked by bin/cleaner to remove/archive jobs \\
config.pl & called by config/Makefile to configure Server \\
dailyreport.p & invoked by bin/report, calls report.pl \\
monitor.tk & invoked by bin/monitor to view logs/database  \\
query.pl & queries master database \\
report.pl  & invoked by bin/report to generate usage reports  \\
restart.pl & invoked by bin/restart to kill and restart server \\
weeklyreport.pl & invoked by bin/weeklyreport, calls report.pl \\
\hline
\end{tabular}
\end{table}

The tables of files throughout Section \ref{implement}
list and briefly describe most of the 
scripts associated with the NEOS Server.  The executables
mentioned in Tables \ref{admin-scripts}, \ref{server-execs}, and
\ref{make-scripts} reside under \file{server/bin/}.  
The scripts in Table \ref{admin-scripts} represent the
guts of the administrative programs found in the \file{bin}
directory and discussed earlier in Section \ref{running}.  
The Server scripts and executables in Table \ref{server-execs}
are dealt with in much detail in Section \ref{submit-flow}.  The scripts 
for Web site construction are further described in 
Section \ref{web-create}.  The
CGI script templates in Table \ref{web-scripts} can be
found under \file{server/lib/cgi/}.  The location of 
these CGI scripts after the Server-specific information 
has been added to the templates depends on the Server
configuration information.  The role of these CGI scripts
in the Web interfaces arises in Section \ref{web-flow}.
None of these scripts should be executed manually, as 
most are called internally by the Server
within a specific shell environment or by the Web server
with specially formatted arguments or input.

\subsection{Submission Flow}
\label{submit-flow}

To explain the flow of submissions through the NEOS Server,
we begin with a holistic approach that briefly mentions the main components
in Table \ref{server-execs}
as part of the process.  We then move on to a more detailed examination of the
various scripts comprised by the Server.

\begin{figure}[htb]
\psfig{file=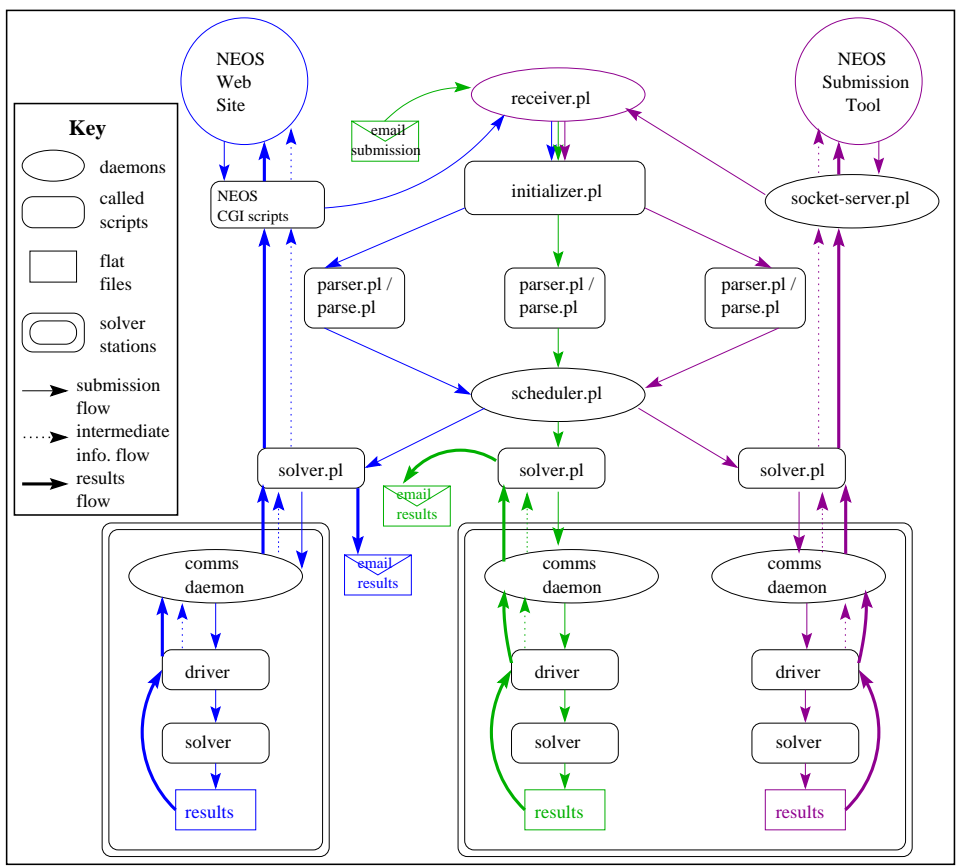, height=4.5in}
\caption{NEOS Server Submission Flow}
\label{Flow} 
\end{figure}

Jobs are processed by the Server in pipeline fashion, beginning 
with the script \script{receiver.pl} 
and ending with the script \script{solver.pl}.  The submission flow becomes 
somewhat complicated by the parallel execution of certain scripts like 
\script{parser.pl}, \script{parse.pl}, and \script{solver.pl}.  Figure
\ref{Flow} shows three simultaneously arriving jobs from each of three
user interfaces.  After being sent in via email, a Web form, or a submission
tool form, the job submissions are located by the \script{receiver}.  
When it sees submissions, the \script{receiver} calls the 
\script{initializer.pl} script to add the submissions to the NEOS jobs
directories.  Then the \script{initializer} spawns \script{parsers} to
decode the submissions.  When the \script{parsers} complete their tasks,
the \script{scheduler} is informed of the waiting jobs.  The \script{scheduler}
makes every effort to refer the jobs to the correct solver in the order they
finished being parsed and spawns a \script{solver.pl\/} script for each 
job in turn.  The
\script{solver} scripts request job execution on the solver station determined
by the \script{scheduler} via a \script{client} script written to handle 
socket communications with the remote NEOS Comms daemons.  When the job is 
complete, results are sent back either through the helper scripts or by
the \script{solver.pl} itself.

While the Web and socket interfaces display a similar 
flow of information, Email submissions require different amounts of 
attention at various stages.  For example,
Figure \ref{Flow} depicts how both the Web and socket submissions are presented
to the \script{receiver} script by NEOS helper scripts (CGI scripts and
the \script{socket server} script).
On the other hand, the \script{receiver} must actively request 
Email submissions
from the email server.  The \script{parser} emails users that their
Email interface submissions have arrived, and the \script{solver.pl} 
script emails
results back to the user through an external mail program (MH) rather than
relying on NEOS helper scripts.  The \script{solver.pl} makes no effort
to transmit intermediate results to the email user as it does to the 
helper scripts; instead it waits until the job is complete to send results.

One distinction between the Web and socket submission interfaces is that the
NEOS-created \script{socket server} daemon runs on the NEOS Server machine 
while the Web server daemon mentioned is the Web server 
(e.g., Apache) that hosts the NEOS Web site.  The Web interface requires CGI 
scripts to facilitate communication with the NEOS Server via files while 
the \script{socket server} written for NEOS contains 
knowledge of the NEOS Server protocols internally.  

Once we have looked at the components of the job submission flow
that are common to all interfaces, we examine the more popular
interfaces in some detail.
Section \ref{socket server} deals with the \script{socket server} both
as a user interface and an interface between the NEOS Server and the
Comms Tool daemons.
The CGI interfaces to the Server are discussed as a part of 
Section \ref{WebServer}.  The following descriptions of Server
scripts are common to all user interfaces except where otherwise
noted.

\begin{table}[htb]
\caption{\textbf{Server Scripts \& Executables}}
\label{server-execs}
\smallskip

\begin{tabular}{|l|l|}
\hline
Executable & Main Purpose \\
\hline
address & extracts email address from all submissions \\
client.pl & code for socket requests\\
counters.pl & solver usage counts/restrictions \& size limit  \\
initializer.pl & creates job directories \& invokes parser.pl \\
parser.pl  & parses jobs into files via parse.pl; signals scheduler.pl \\
parse.pl  & performs the actual parsing using token config. files  \\
receiver.pl & looks for incoming jobs and invokes initializer  \\
scheduler.pl & handles job queues; forks solver.pl's \\
socket-server.pl & handles socket requests to Server \\
solver.pl & requests Comms daemons execute job, using client.pl \\
\hline
\end{tabular}
\end{table}

\subsubsection{The Receiver}

The script \script{receiver.pl} is one of the three Server 
daemon processes that should
always be running, and the job of the \script{receiver} is to check 
for incoming jobs from any one of the interfaces in its 
spooling directory.  

Submission tool jobs are initially written to the directory 
\file{\var{server\_var}/spool/\linebreak[0]SOCKET}, Kestrel jobs
are found in \file{\var{server\_var}/\linebreak[0]spool/\linebreak[0]KESTREL},
and, similarly, WWW submissions are deposited in the 
directory \file{\var{server\_var}/\linebreak[0]spool/WEB}.  
When one of the NEOS Server interfaces is writing a job submission file
to its spooling directory, it must first choose a unique file name, 
which takes the general form \file{\var{interface}.\var{unique\_filename}}.
In order to ensure that all job files are completely 
written before being collected by the \script{initializer}, 
the \script{socket server},
\script{nph-solver-*.cgi} scripts, and the Kestrel server will touch
the file \file{DONE.\var{interface}.\var{unique\_filename}} when finished
writing the submission file and then signal a job arrival by touching the 
file \file{MAIL} in the interface's spooling directory.  
The original NEOS Server implementation checked for Email jobs in much 
the same way as other jobs by looking for a nonempty 
\file{/var/spool/\linebreak[0]mail/\var{username\/}} file.  
Since checking mail via the network file system has fallen out 
of favor, the current \script{receiver} script queries 
the POP3 mail server for new mail. 

The \script{receiver} checks for \file{MAIL} files every second
and queries the mail server only after a number of unsuccessful
rounds through the spooling directories so as not to overwhelm the mail server.
It launches the \script{initializer} as a blocking call
if there are submissions waiting, with the argument \command{-mail} 
if there is new mail waiting on the mail server.

\subsubsection{The Initializer}

The script \script{initializer.pl} is invoked by the receiver 
once jobs have arrived
from any one of the interfaces.  The primary job of the 
initializer is then to

\begin{itemize}

\item{download email from the mail server, and refile messages in the \file{\var{server\_var}/tmp} directory if the \command{-mail} argument is present;} 
\item{throw out any junk mail;}
\item{assign a job number to each legitimate submission;}
\item{create the job directory 
\file{\var{server\_var}/jobs/job.}\var{number\/}, 
      where \var{number\/} is the number assigned to the 
job by the \script{initializer};}
\item{move the job from its spooling or temporary directory (i.e., 
\file{\var{server\_var}/tmp}, 
      \file{\var{server\_var}/\linebreak[0]spool/\linebreak[0]SOCKET},
\file{\var{server\_var}/\linebreak[0]spool/\linebreak[0]WEB}, or 
\file{\var{server\_var}/\linebreak[0]spool/\linebreak[0]KESTREL}) to the file name
       \file{job.received} in its new job directory; and}
\item{fork the script \script{parser.pl} to parse the 
contents of \file{job.received}.} 
\end{itemize}

The \script{initializer}'s first order of duty in each spooling
directory involves checking for and deleting the \file{MAIL} file.
If \file{MAIL} is present, the script looks for a matching 
\file{DONE} file for every submission and ignores those files lacking a
match.  The \script{initializer} assigns a job number based on
the contents of \file{\var{server\_var}/lib/next\_serial\_number}, which
it increments, and outputs a file
for later input by the Web, Kestrel, and socket servers.  The job
 number written to this serves as a handle to these interfaces
to obtain intermediate output from a solver as well as final 
job results.  

While users of the Email interface do not normally 
receive intermediate results, once the \script{parser} has found a 
return address, it sends reply mail with the
job number and password so that users can check the status of their
jobs on the Web or refer to the job number with questions about 
particular submissions.  The Server script 
\script{solver.pl} assumes responsibility for mailing final results 
back to the user.  The \script{solver} script lies a few steps 
ahead of us, though.  First, we focus on the other duties of the \script{parser}.

\subsubsection{The Parser}

Parallel execution in the Server begins with the script \script{parser.pl}, 
as the initializer forks
a new Perl process for each job. The primary role of the \script{parser} 
is to read in the file 
\file{job.received} and decompose the job into its constituent parts:

\begin{itemize} 
\item{Address of the sender}
\item{Solver requested}
\item{Data to be input to the solver}
\end{itemize} 

Once the sender's IP or email or address has been determined 
(via the \script{address} executable examining the ``\verb+From:+'' 
line in \file{job.received}),
this information is saved in the file \file{job.address}.  
Because both socket and WWW submissions will yield i.p. addresses, 
the WWW submission addresses are tagged with \texttt{WEB\_USER}; 
and the addresses of sample WWW submissions have \texttt{TRIAL\_WEB\_USER} 
preappended to them.  The solver requested is saved in the
file \file{job.type}.  The \script{parser.pl} script then 
calls on \script{parse.pl} to compare the data in the remainder 
of \file{job.received} with the information in the requested solver's 
token configuration file and map the data to files accordingly.  The NEOS Server
interfaces that token delimit files for the user can make parsing
somewhat simpler by omitting the end token and giving the begin 
token as \command{begin.token[\var{size}]},
where size is the number of bytes of content for the file, beginning on the
next line.  The \script{parse.pl} 
executable also checks whether any of the files submitted were in compressed 
format and, if possible, decompresses them.  The \script{parse.pl} script further 
massages the data so that input from DOS or Mac operating
systems resembles that expected by a Unix-based solver.  The \script{parse.pl} 
script excludes binary files, which are flagged by a begin token ending 
in \emph{BINARY}, from this operating system format conversion.  The correct
parsing of binary files relies on having the \var{size} available.

When \script{parse.pl} has finished its tasks, the parser is ready to 
request scheduling of the job for execution.  The \script{scheduler} has its own 
spooling directory in \file{\var{server\_var}/spool/} (similar
to those checked by the \script{receiver}).  To wake up the \script{scheduler} 
daemon, the \script{parser} saves the job type into the file 
\file{job.\var{number}} in the directory \file{\var{server\_var}/spool/JOBS/} and 
signals the \script{scheduler} by creating 
the empty file \file{\var{server\_var}/spool/JOBS/JOBS}.

\subsubsection{The Scheduler}

The \script{scheduler.pl} script is the second of the three 
Server daemons, and its primary goal is to 
fork a \script{solver.pl} script for each job.  The 
\script{solver} script is then responsible for 
connecting to the NEOS Comms Tool daemon
running on the machine of the remote solver.  
Jobs that cannot be scheduled on their requested 
solver are also handled by the \script{solver} script on the 
NEOS Server host machine, which will generally return job 
results explaining why the job could not be scheduled.

The important information needed by the \script{scheduler} 
for each job is the job number, the job type 
and solver identifier, the machines (and port numbers) running the 
solver's NEOS Comms Tool daemon,
and the jobs already executing on each solver station.  
The \script{scheduler} internally stores a list of all 
job numbers, their solver, and their state (e.g., executing, queued 
for execution, scheduling error).  The \script{scheduler} also
tracks the current station use for each solver (e.g., solver \solver{gams:bdmlp}
has 3 jobs running on its \file{fire.mcs.anl.gov} station).

The \script{scheduler} executes only as many jobs of one type 
on a solver station as the solver administrator indicates are 
allowed on that station when registering the solver.  
The machines that are currently executing jobs
are represented by the files in the directory 
\file{\var{server\_var}/proc/stations},
 where each file represents a solver on a 
particular machine.  For example, the file 
\file{fire.mcs.anl.gov-GAMS:BDMLP} is created by the \script{scheduler}
to show that the \solver{gams:bdmlp} solver is currently running on 
the machine fire.mcs.anl.gov.  If the file is nonempty, 
it will contain an integer
representing the number of \solver{gams:bdmlp} jobs running on \file{fire}.
It is the responsibility of the \script{scheduler} to decrement or 
remove this file when the forked 
\script{solver} script has completed execution.  

The file \file{\var{server\_var}/lib/station\_type\_port} is maintained by the 
\script{scheduler} and stores a list of all solver machines, including
which port the NEOS Communication Packages are listening on.  The 
\script{scheduler} will consult this list before
forking a \script{solver} script in order to direct it to a particular machine and port.  For example, the machine/port
entries in \file{\var{server\_var}/lib/station\_type\_port} 
for the \solver{gams:bdmlp} solver might look like
\medskip

\noindent
\command{fire.mcs.anl.gov:GAMS:BDMLP:4001\\lava.mcs.anl.gov:GAMS:BDMLP:4000\\ember.mcs.anl.gov:GAMS:BDMLP:4007} ,
\smallskip

\noindent
which gives the \script{scheduler} three machines to choose from when executing a job. 

\subsubsection{The Solver and Client}
\label{solver-script}

The \script{solver.pl} script is responsible for initiating the data 
exchange between the NEOS Server and remote solver stations for
 all job submissions.
  In the event that a job cannot be scheduled on a remote machine, 
the \script{solver} script writes a message to the results file
and mails users of the Email interface.
The \script{solver} script requests that legitimate job 
submissions be executed on their solver stations.  If the solver
station is \emph{localhost}, as is the case for the \solver{admin}
solvers, \script{solver.pl} finds and executes the solver driver
itself, as discussed in Section \ref{admin-imp}.  In order to 
execute jobs on remote machines, the NEOS Server must act as the client 
making requests to the NEOS Comms Tool daemon as the server.  For these 
submissions
 \script{solver.pl} calls the \script{client.pl} script, which actually 
connects to the NEOS Comms Tool daemon on the remote machine.  

The 
\script{solver}
 script relies on \script{client.pl} to pass on requests to the NEOS Comms 
Tool daemon to accept uploaded user data and begin execution of the 
remote solver 
driver.  The \script{solver} script invokes the \script{client} script with 
all of its output redirected to the file \file{job.out} so that the standard
 output from the solver driver being piped by the NEOS Comms Tool 
daemon and any error messages from the NEOS Comms Tool daemon can be 
displayed to the user by the 
socket and Web servers.  When the job is complete, the \script{solver} script 
forwards results to Email users and WWW interface users who request email.  

The \script{solver} scripts also keep the \file{master} database.  When
results are in, \script{solver.pl} writes a 256-byte descriptive 
entry into the \file{master} database at the position in the 
database corresponding to the job 
number.  Then \script{solver.pl} signals the waiting WWW and socket servers 
that the job is finished by writing an end token to the \file{job.out} file, 
and it creates the empty file \file{DONE} in the job's working directory to 
mark the job as completed.
Because job completions will not always occur in the order in which the 
jobs are received, the master database will at times contain sections of 
garbage. The \script{report}
and \script{cleaner} scripts replace the unreadable sections of the master
database with white space so that the Server administrator can query
the database as part of Server monitoring.

\subsection{The Socket Server}
\label{socket server}

The \script{socket server} carries the responsibility for responding 
to all requests from the socket submission interfaces and solvers 
(via the NEOS Comms  Tool daemon). 

In general, a TCP/IP server listens for and accepts socket connections 
with arbitrary clients. After being
connected, the \script{socket server}  reads the client's request 
from the socket, processes the request, 
and sends a response back over the same socket connection.  The Perl 
TCP/IP client 
\file{server/bin/}\script{client.pl} is distributed with the socket 
submission tools and the Comms Tool to make requests to the NEOS Server 
that the \script{socket-server.pl} script can fulfill. 

\begin{figure}[hbt]
\centerline {\psfig{file=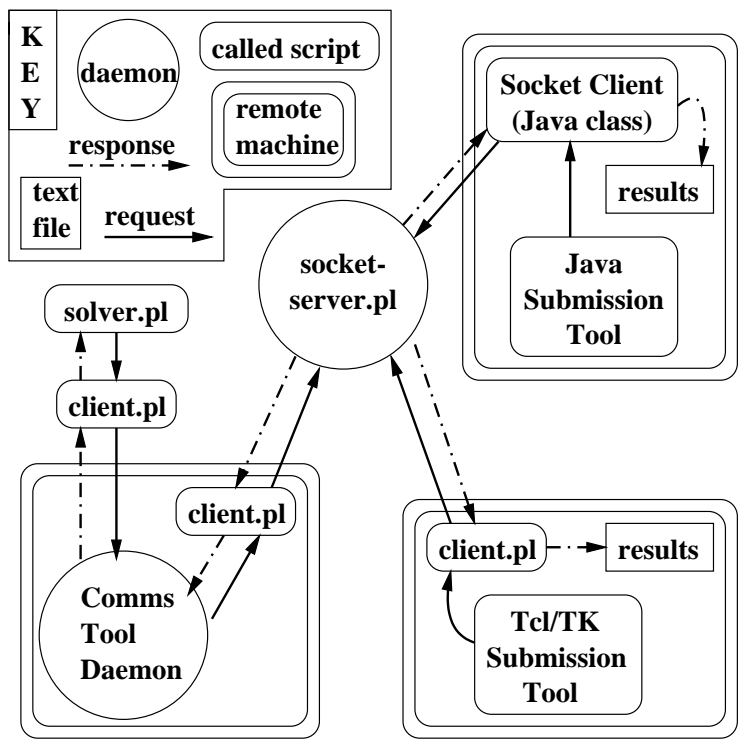}}
\caption{NEOS Socket Communication}
\label{socket-server} 
\end{figure}

\subsubsection{Interface with the Job Pipeline}

When the \script{socket server} receives a job from a client, it must submit 
this job to the Server on behalf of the client.  The entry point for 
these jobs is the spooling directory
\file{\var{server\_var}/spool/SOCKET}.  In this directory the 
\script{socket server}
 writes job submissions with a header.  In other words, all submissions look 
 similar to email submissions in that they start
with a ``From:'' line.   The submission tool at the user end
actually composes the job submission, and the \script{socket server} 
just reads in the data over the TCP/IP connection. Then the 
\script{socket server} creates a file 
in \file{\var{server\_var}/spool/SOCKET} under the
name \file{job.}\var{number\/}, where \var{number\/} is an internal 
variable in the \script{socket server} used to track incoming jobs.

After \file{job.123}, for example, has been written, the 
\script{socket server} signals the \script{receiver}
by creating the empty file \file{\var{server\_var}/spool/SOCKET/MAIL}.  
At this point the \script{receiver} wakes
up, and the job is processed by the \script{initializer}.

The \script{socket server} will know that its job has been accepted once 
the file \file{job.123} 
has been renamed to \file{job.123.number}. This new file will 
contain the job number assigned by the \script{initializer}.
This job number will need to be returned to the client that 
originally submitted the job because it will be 
used as a handle when later downloading the job results.

\subsubsection{Socket Communications Requests}
\label{socket-requests}

The sections below discuss the requests accepted by the \script{socket server},
 how it processes each request, and the response sent back over the 
socket to the client.  There are two categories of requests: those issued 
by the \script{client} script or Java \script{SocketClient} on 
behalf of a submission tool and 
those issued by \script{client.pl} on behalf of a Comms Tool daemon.

The first line of each request to the \script{socket server} represents the 
userid of the client, and the second line represents the actual request.  
After the userid has been read by the \script{socket server}, the machine 
name is determined (as per the TCP/IP protocol), and the request is then read.
  The userid and machine name are used for logging purposes so that
the logs display all requests to the socket server.   

All requests can be tested by executing \script{client.pl} directly using the
\command{perl} command with the arguments
\medskip

\noindent\command{client.pl -server <hostname>:<port> -request <server request>} ,
\medskip

\noindent where \command{<hostname>} represents the machine name of 
the server (e.g. \command{opt.\linebreak[0]mcs.\linebreak[0]anl.gov}), 
\command{<port>} represents the port the \script{socket server} is 
running on (e.g. \command{3333}), and 
\command{<server request>} represents the actual request 
(e.g. \command{verify}).  The \script{client.pl} 
script is coded to automatically 
send the userid of the requester.

Unless otherwise stated, the first line of server output 
 in response to a request represents the amount of data
to be transferred back to the \script{client}.  All lines thereafter 
represent the output of the request.

\subsubsection{NEOS Submission Tool Requests}

The requests in this section describe the protocol between the 
NEOS server and the NEOS submission tools.  

\paragraph{solver\_list}

\noindent
The request \command{solver\_list} returns to the client a list of all the 
registered solvers. Each line of output contains information of the 
form 

\smallskip
\noindent
\var{Solver Name}=\solver{type:solverid}
\smallskip

\noindent
The following line of output, for example, would be transferred for the \var{TRON (AMPL input)\/} solver whose type and id is \solver{bco:tron-ampl}:

\begin{verbatim}
TRON (AMPL input)=BCO:TRON-AMPL
\end{verbatim}

\paragraph{admin\_list}

\noindent
The request \command{admin\_list} returns to the client a list of all administrative solvers 
in addition to all registered solvers.  The output of this request is identical to that
generated by \command{solver\_list} except that the administrative solvers are included, of 
which there are currently only

\begin{verbatim}
Adding/Modifying a Solver=ADMIN:ADDSOLVER
Enabling/Disabling a Solver=ADMIN:STATUS
Kill Job=ADMIN:KILL_JOB
\end{verbatim}

\paragraph{help\solver{type:solverid}}

\noindent
The request \command{help}\/\solver{type:solverid} returns to the client 
the submission tool help file for the requested solver.  

\paragraph{config\/\solver{type:solverid}}

\noindent
The request \command{config}\/\solver{type:solverid} returns to the client the token configuration file for the
requested solver.  

\paragraph{begin job \var{size}}

\noindent
The request \command{begin job} \var{size} is used to submit jobs 
to the server, where \var{size} 
is the size in bytes of the job being submitted.  After the job has 
been submitted, the first
line of output to the client is the job number, and all remaining 
lines represents the standard
output/error from the executing job.  These lines are transmitted 
as they are generated by the 
solver.  There is no line specifying the size of the transfer from 
the server, as this cannot be predetermined while the solver is executing. 

The following example can be followed to submit a job to the server 
specifying \command{help}
for the administrative solver \solver{admin:addsolver}. 

Begin by executing the command

\begin{verbatim}
perl server/bin/client.pl -server opt.mcs.anl.gov:3333 \
      -request "begin job 33"
\end{verbatim}

After the connection has been established you should see output similar to

\begin{verbatim}
client.pl: connecting to opt.mcs.anl.gov:3333
client.pl: connected
client.pl: sending request
\end{verbatim}

Type in the following lines as input to the executing script, followed by
an end signal, usually \command{CTRL-D}:

\begin{verbatim}
type admin
solver addsolver
help
\end{verbatim}

And the remaining output from the server should be similar to

\begin{verbatim}
client.pl: 33 bytes sent
client.pl: receiving data
7773
Welcome to NEOS!
<CLEAR_SCREEN>
Parsing:
        0 bytes written to help.type ()
Welcome to NEOS!
<CLEAR_SCREEN>
Scheduling:
  You are job #7773.
  Solver Queues:
    ADMIN:HELP: 7773:
  Jobs Executing:
    No jobs executing.
<CLEAR_SCREEN>
Parsing:
               <END_STANDARD_OUT>
        0 bytes written to help.type ()
client.pl: 339 bytes received
client.pl: exiting
\end{verbatim}

Ignore for now all of the marked-up commands, such as \command{<CLEAR\_SCREEN>}, as these
are directives to the submission tools and Web CGI scripts to do exactly that, clear
the screen for the user. What is important to note is the job number that was returned
as the first line of output, \command{7773} in this example. Following this job number is the 
standard output/error from the running job. This is the same information that is written 
into the file \file{job.out} in the job's temporary working directory in \file{\var{server\_var}/jobs}. 

\paragraph{get results \var{number}}

\noindent
After a job has completed, the submission tool retrieves the results using
the request \command{get results} \var{number}, where \var{number} is the 
number of the job and should have been returned as the first line of output when
the job was submitted with the request \command{begin job}. 

In the preceding section we were assigned a job number of 7773. You can use \script{client.pl}
to send this request for you:  

\begin{verbatim}
perl server/bin/client.pl -server opt.mcs.anl.gov:3333  \
      -file myResults -request "get results 7773"
\end{verbatim}

\noindent
And you should see output similar to the following:

\begin{verbatim}
client.pl: connecting to opt.mcs.anl.gov:3333
client.pl: connected
client.pl: sending request
client.pl: 17 bytes sent
client.pl: receiving data....
client.pl: 26348 bytes received
client.pl: exiting
\end{verbatim}

To see the actual job results, you will need to view the file \file{myResults}. Without the
\command{-file} option the results would have been displayed directly to the console. 

\paragraph{submit-\var{version}.tk}

\noindent
The request \command{submit-}\var{version}\command{.tk} is used to 
retrieve the Tcl/\linebreak[0]Tk code for
the Unix submission tool.  This is the code responsible for generating 
the submission tool
GUI.  The version number is used to distinguish between different 
versions in the event that
future Tcl/\linebreak[0]Tk submission tool distributions require different GUIs.  
Currently, the only version of 
the Tcl/\linebreak[0]TK code is \command{submit-1.0.tk}.  

\paragraph{java-client-\var{version}.jar}

\noindent
The request \command{java-client-}\var{version}\command{.jar} returns the 
Java archive file including all of the java Submit Client class files.  If the
\var{version} requested is the same as the most recent version of 
the Submit Client, then the user's version is already up to date; and
the server returns a message to that effect.  Otherwise, the latest version
is returned.

\paragraph{submit-main.txt}

\noindent
The request \command{submit-main.txt} returns to the client the ``homepage'' for the NEOS submission
tools.  This text is displayed in the main window of the submission tool when it is first started. 

\paragraph{submit-help.txt}

\noindent
The request \command{submit-help.txt} returns to the client the help file for the NEOS submission
tools.  This text is displayed in the main window of the submission tool when the Help button
is clicked. 

\paragraph{verify}

\noindent
The request \command{verify} may be used by NEOS submission 
tools to verify the identify 
of the NEOS Server. The output of this request is the name of the server. 

\subsubsection{Comms Tool Requests}
\label{neos-requests}
The requests discussed in this section are used by the
Comms Tool daemons to
coordinate all communication between a solver and the Server. 
The primary responsibilities of the 
Comms Tool are to accept connections from the Server, download user data, 
execute the solver, and return the job results.

Just as with the Tcl/\linebreak[0]TK submission tool, the Perl script 
\script{client.pl} is distributed with the 
Comms Tool and is used to issue all requests to the Server.

\paragraph{verify}

\noindent
The request \command{verify} may be used by Comms Tool to verify the identify 
of the Server. The output of this request is the name of the Server. 

\paragraph{begin results \var{size number}}

\noindent
The request \command{begin results} \var{size number} is used 
to return job results to
the Server.  The arguments represent the size of the job results 
in bytes and the job number.  

\paragraph{comms-\var{version}.tk}

\noindent
The request \command{comms-}\var{version}\command{.tk} is similar to \command{submit-}\var{version}\command{.tk}
in that it downloads the Tcl/\linebreak[0]Tk code used to create the Comms Tool GUI.  The version
number is used to distinguish between different version in the event that later distributions
of the Comms Tool require a different GUI.  Currently, the only version is 1.0.

\paragraph{comms-daemon-\var{version}.pl}

\noindent
The request \command{comms-daemon-}\var{version}\command{.pl} 
returns the Unix Perl code for the main 
communications daemon that accepts incoming connections 
from the server. Currently, the only version
is 1.0.

\paragraph{comms-backup-\var{version}.pl}

\noindent
The request \command{comms-backup-}\var{version}\command{.pl} returns the Perl code responsible for
editing the crontab file in the event that the crontab entry for the communications daemon
needs to be enabled. Currently, the only version is 1.0.

\paragraph{disable-backup-\var{version}.pl}

\noindent
The request \command{disable-backup-}\var{version}\command{.pl} returns the perl code responsible for
editing the crontab file in the event that the crontab entry for the communications daemon
needs to be removed. Currently, the only version is 1.0.

\paragraph{register \var{solver password port}}

\noindent
The request \command{register} \var{solver password port} is issued by \script{client.pl} on behalf of a solver's
communication daemon.  This request tells the NEOS Server what port the communications daemon is 
listening on.  This request is sent to the Server whenever a communications daemon is restarted.
\subsubsection{Server Requests to the Comms Tool}
\label{comms-requests}
This section discusses the requests accepted from the NEOS Server by 
the Comms Tool daemon (\script{comms-daemon-\var{version}.pl}). 

\paragraph{verify}

\noindent
The request \command{verify} is made by the Server 
to verify the identity of a communications
daemon. The output returned to the Server from the 
communications daemon is of the form 
\solver{type:solverid}\command{@hostname:port}.

\paragraph{begin job size number}

\noindent
The request \command{begin job} \var{size number} is made by the 
Server to remotely execute
a particular solver, where \var{size} represents the size of the user 
data to be transferred
to the Comms Tool and \var{number} represents the job number. This job 
number is later used in the \command{begin results} request 
after the application exits  and the Comms Tool daemon 
returns the results to the server.

The user data uploaded to the Comms Tool represents the contents of a
 file created with \command{tar} and compressed with \command{gzip}.  
The Comms Tool must uncompress
and untar the data in order to present the user's data 
files to the solver.  When the solver driver is
executed, the data has already been untarred and 
uncompressed so that the files are in the 
current working directory.

\paragraph{kill job number}

\noindent
The request \command{kill job} \var{number} may be made by the NEOS Server
through the \solver{admin:kill\_\linebreak[0]job} solver.  If the \script{comms-daemon}
script running on the remote solver station's list of running jobs
includes the \var{number} requested and the solver's jobs have been
labeled ``killable'' by the solver administrator when starting the Comms
Tool daemon, then the \script{comms-daemon} will kill the entire process
group associated with that job.  Solver administrators should test the
\solver{admin:kill\_job} solver's behavior with their solver to make sure
that killing the process group does effectively kill all descendants of
the child job.  Some Unix systems handle process groups differently from
others.  Also, some application software may reset the process group id
for its own ends or run some processes on other machines, so 
\solver{admin:kill\_job} may not work for all solvers.  In these cases, the
Comms Tool daemons for the solver should simply be started with the 
``killable'' option deselected.

\pagebreak\section{Administrative Solver Implementation}
\label{admin-imp}

Four solvers currently come packaged with the NEOS Server 3.0.
These solvers, introduced in Section \ref{admin-solvers},
are identified as \solver{addsolver}, \solver{help}, \solver{kill\_\linebreak[0]job},
and \solver{status}.
The registration information for these solvers can be found 
under the directories 
\file{server/lib/\linebreak[0]admin\_solvers/\linebreak[0]ADMIN:\var{identifier}/}, stored in the same format the 
\solver{admin:addsolver} uses for other solvers.  The 
source code for the administrative solvers is also present
in these directories, so that it can be executed directly
on the NEOS Server host machine.  Because they are integrated
with the Server and run under the Server userid, 
\solver{admin} solvers may call on Server
scripts to complete their tasks or write to Server directories.
Because their registration information is stored similarly,
the interfaces of the administrative solvers are constructed
in the same way as for other solvers, allowing the same 
familiarity afforded by the automated addition of other solvers to
the NEOS Server pool.

When a job is submitted to an administrative solver, the 
submission arrives on the Server in the appropriate 
\file{\var{server\_var}/spool/} directory
according to the interface of origin.  The \script{parser}
script is the first part of the submission flow that 
distinguishes an \solver{admin} job from other jobs.  The
parser allows users greater flexibility in invoking the
\solver{admin:help} solver by sending any submission with
\emph{help} in the first nonblank line after the header
to solver \solver{admin:help}, regardless of other parsing
oddities.  The \script{scheduler} also makes special dispensation
for \solver{admin} jobs by automatically assigning them 
\emph{localhost} as their solver station and by 
redirecting the search for their solver registration information
to the directory \file{server/lib/admin\_solvers/}.  The
\script{solver.pl} script assigned to an \solver{admin} job
makes no effort to contact a Comms Tool daemon through the
\script{client.pl} script.  Instead, the \script{solver} reads the name 
of the driver for each \solver{admin}
solver from \file{ADMIN:\var{identifier}/SOLVE} and 
invokes the administrative solver driver itself.
With these relatively minor adjustments,
\solver{admin} jobs can be submitted and processed like any others.

The \emph{Adding/Modifying a Solver} administrative solver 
comprises a group of Perl scripts for checking registration
information and adding solvers to the NEOS Server.  If each
piece of data sent in by solver administrators is in the
correct format, the \solver{addsolver} ensures that there is
a directory for the solver under \file{\var{server\_var}/lib/solvers/}.
In the case of a registration update, old solver stations in the
file \file{\var{server\_var}/\linebreak[0]lib/solvers/\linebreak[0]\var{type}:\linebreak[0]\var{identifier}/\linebreak[0]STATIONS}
not in 
the new list are slated for deletion by the \script{scheduler}
from its list of available solver stations and the ports bound by their
Comms Tool daemons.  The \solver{addsolver} copies the 
new list of stations and other information to the correct
\file{\var{server\_var}/lib/solvers/\var{type}:\var{identifier}/} directory.
The \solver{addsolver} adds or updates the solver's entry in the
\file{\var{server\_var}/lib/solver\_list}.  Then it calls 
\file{server/bin/\script{make-solver.pl}} with the correct argument
to build the Web pages and CGI scripts for the Web interface to 
the solver.  When the process is finished, solvers are registered as
available to all interfaces enabled for the Server.

The \emph{Enabling/Disabling a Solver} administrative solver,
\solver{status}, changes the availability of a registered
solver.  Depending on the data sent in with a \solver{status}
job submission, \solver{status} may just disable a solver
by removing it from the \file{\var{server\_var}/lib/solver\_list} 
and calling on
\file{server/bin/\script{make-server-solvers.pl}} to update
the Web page listing of solvers.  The reverse operations
may be performed at any time to enable such a solver again.
If the solver administrator requests deletion,  the
solver is first disabled.  Then the solver registration information
is deleted from under \file{\var{server\_var}/lib/solvers/},
the Web pages and scripts are deleted, and the solver 
is removed from the \script{scheduler}'s list of stations and
ports.  The solver can be restored only by reregistering it
through \solver{admin:addsolver}.

The \emph{Help} solver can give help relating to a particular
solver or for the NEOS Server in general.  The \emph{Help}
solver provides individual solver help aimed at the Email
interface only. The solver help files for the Web interface are
built into the Web site, and the solver submission tool help files
are available by request to the \script{socket server}.
The \solver{admin:help} solver searches the library directory
of the solver requested for the \file{email-help} file.  If
present, \solver{help} returns this text to the user along with
information on which interfaces are currently enabled for the
NEOS Server.  If the solver administrator fails to provide an
email help file, \solver{help} informs the user that no help is 
available.  If the user does not request a particular solver or
the help submission cannot be parsed, \emph{Help} returns
a  \file{general\_help} text from its own directory along
with a list of all registered solvers, including administrative
solvers.  It also provides further tips on how to use the
\solver{admin:help} facility.

The \emph{Kill Job} administrative solver relates to specific
jobs.  If the requested job number's directory still exists 
under \file{\var{server\_var}/jobs/} and \solver{admin:kill\_job}
can find the job's solver station and port, then \emph{Kill Job} 
sends a socket request through the Server's 
\file{server/bin/\script{client.pl}} 
script to the Comms Tool daemon associated with that solver and station.
\emph{Help} returns the output from the request to the user,
which usually states that the job is dead, the \solver{kill\_job}
solver is not enabled by that Comms Tool daemon, or that
the daemon has no record of the job.  The correct implementation of
the \solver{kill\_job} request by the Comms Tool daemon is hence
not required, because it would be impossible on some systems.  
Solver administrators are expected to test the feature on their
site before enabling it.  The administrative \emph{Kill Job}
solver performs an attempt at its task, regardless.

As contemplate new features for the NEOS Server, we may
decide to implement new administrative solvers.  The task
is not too difficult for any Server administrator.  If the
format of the existing administrative solvers is followed 
and the new solver added to the \file{server/lib/admin\_list},
then the administrative solver becomes available with a simple
rebuild of the Server.

\pagebreak\section{Web Site/Interface Implementation}
\label{WebServer}
If a Web site is not desired or cannot be supported, the NEOS Server will
still function fully via its socket and Email interfaces.  We recommend a
Web presence for advertising and educational purposes, however, and note
that, for the Optimization Server, this is our most popular job submission 
interface.

Many of the CGI scripts in Table \ref{web-scripts} support miscellaneous
user requests that can be made throughout the Web site.
Whenever a user clicks on a NEOS Web 
form's \command{Submit} button, the Web server that hosts the page 
executes a NEOS CGI script.  For example, when users send in comments
from either the NEOS Server comments page or one of the individual solver
comments pages, \script{server-comment.cgi} or \script{solver-comment.cgi}
handle emailing the comments to the appropriate addresses.  When users
submit their email address for the \emph{neos-news} list, 
\script{list.cgi} forwards their request to the list server.  The addresses
of people downloading various NEOS packages are recorded by the
\script{download} scripts.  The other files in Table \ref{web-scripts}
play a more active role in the job submission flow as it relates specifically
to the Web interfaces.

\begin{table}[htb]
\caption{\textbf{Scripts and Executables for the Web Server}}
\label{web-scripts}
\smallskip

\begin{tabular}{|l|l|}
\hline
General Template & Main Purpose \\
\hline
cgi-lib.pl & parses Web uploads; sets upload limit  \\
check-pwd.cgi & verifies check-status password;\\
& outputs job results to Web browser\\
check-status.cgi & gateway to job results \\
downloads.cgi & logs emails of Tcl/TK tool downloaders\\
java-downloads.cgi & logs emails of Java tool downloaders\\
list.cgi & adds email addresses to neos-news mailing list\\
server-comment.cgi & emails user comments to Server administrator\\
server-downloads.cgi & logs emails of NEOS-4.0 downloaders\\  
tempfile.cgi & creates temporary, unique files names \\
\hline
\end{tabular}
\bigskip

\noindent
\begin{tabular}{|l|l|}
\hline
Solver Script Template & Main Purpose \\
\hline
nph-solver-sample.cgi & handles sample submissions\\
nph-solver-www.cgi & handles Web interface submissions\\
solver-comment.cgi& emails user comments to solver administrator\\
\hline
\end{tabular}
\end{table}

\subsection{The Web Server Role in Submission Flow}
\label{web-flow}

The NEOS Server acts to take full advantage of Internet technology by
creating a potentially extensive Web site.  In addition to the Server homepage,
solver listing, and various informational pages, the NEOS Server
creates a homepage for each solver added.  From these solver
homepages, users can access the NEOS Server's Web interfaces for
job submission.

The NEOS Server works hand in hand with a Web server to orchestrate
Web interface submissions.  
The \script{nph-solver-www.cgi} script, in conjunction with 
the \script{cgi-lib.pl}
routines, uploads a user's files and other data entries.  By referring to
the solver's token configuration file, it 
composes one submission file in the NEOS Server's token delimited format.  
Once the submission file has 
been created and given a unique file name by \script{tempfile.cgi}, 
\script{nph-solver-www.cgi} leaves 
a signal of the pending submission for the 
\script{receiver} in the form of an empty \file{MAIL} file under 
\file{\var{server\_var}/spool/WEB/}.  The \script{initializer} facilitates 
communication for Web interface users by writing the job number assigned
to the submission to a file the \script{nph-solver-www.cgi} 
script expects to find.
Once \script{nph-solver-www.cgi} has the job number, it can read in
the intermediate information being written to 
\file{\var{server\_var}/job/job.\var{number}/job.out} by \script{solver.pl} 
and write it in server-push HTML format.  The tag, 
\command{<END\_STANDARD\_OUT>}, that \script{solver.pl} writes to 
\file{job.out} tells \script{nph-solver-www.cgi} that the job results are 
available.  CGI script \script{nph-solver-www.cgi} then formats the results 
as HTML text and prints a redirection tag from the intermediate output to
the final results.  

The solver's \command{Sample Submission} interfaces
create similar output through the \script{nph-solver-sample.cgi}
script as users would see from the \script{nph-solver-www.cgi} script, 
but \script{nph-solver-sample.cgi} expects sample submissions 
in token-delimited format.  Instead of uploading user files and
inserting the solver tokens, \script{nph-solver-sample.cgi} downloads
the selected Web-accessible sample file registered by the solver administrator
from its URL and signals the Server of
a waiting Web submission through the \file{MAIL} file.  
The CGI scripts' method of interacting with the NEOS Server through
files necessitates that the Web server have access to the file system
where the NEOS Server's \var{server\_var} directory is mounted.  The
two servers need not reside on the same machine, but they must share
certain files to work together.

From the user perspective, the Web forms offer a straightforward,
probably familiar interface.  The users submit information via a Web
form, watch the intermediate information about the solution stages 
grow as the Web server sends available data, and finally view the job's 
results.
Because pushing results from a Web server to a browser over an extremely
lengthy time (e.g. when a job runs for hours) is prone to network failure, 
users
are given job number and password information at the top of the intermediate 
results page as well as the URL of the \script{check-status.cgi} script.
Through the Web page created by \script{check-status.cgi}, users are able
to submit their job number and password and request the latest 
intermediate results or the job's final results.  The \script{check-pwd.cgi}
script verifies the job password and retrieves text directly from the 
\file{\var{server\_var}/jobs/job.\var{number}} files.  
Because \script{check-pwd.cgi}
finds its information from the job directory, users who have 
sent in their submissions to the Server through other interfaces
 can also retrieve job information through the Web.

\subsection{Web Site Creation}
\label{web-create}

The NEOS Server builds its core Web site using the configuration information
provided by the Server administrator.  The NEOS Server requires such critical
information as the name of a Web-accessible parent directory where
it can write its HTML files and a similar location for CGI scripts that
can be executed by the Web server.  The Server administrator also needs
to know the URLs for these directories before NEOS can build its site so that
the site can be fully interconnected.  The scripts responsible for creating
the Web site reside under \file{server/bin/}, and their names generally begin 
with \script{make}, as shown in Table \ref{make-scripts}.  Once the 
Server configuration script \script{config.pl} has collected all of the
necessary information during the interactive \command{make} process
described in Section \ref{configure}, 
\script{config.pl} calls on \script{make-all.pl} to build the Web site.

\begin{table}[!htb]
\caption{\textbf{Scripts to Generate HTML \& CGI}}
\label{make-scripts}
\smallskip

\begin{tabular}{|l|l|}
\hline
``make-'' Prefixed Script & Main Purpose \\
\hline
all.pl & builds misc. pages; calls other make-'s \\                 
server-homepage.pl      & makes server's homepage \\    
server-solvers.pl       & makes solvers Web page\\
solver.pl & calls other make-solver-* scripts \\
solver-comment.pl & makes solver's comment form \\    
solver-comment-cgi.pl  & completes solver's comment CGI script \\
solver-email.pl  & completes solver's template HTML page   \\
solver-homepage.pl & makes solver's homepage on server \\
solver-sample.pl & builds solver's sample submission form \\
solver-sample-cgi.pl & completes sample submission CGI script \\  
solver-template.pl & builds solver's email template  \\
solver-www.pl & builds solver's Web interface form \\
solver-www-cgi.pl & completes Web interface CGI script \\
\hline
\end{tabular}
\end{table}

The script \script{make-all.pl} contains a routine for taking
the template HTML files under \file{server/\linebreak[0]lib/html/} 
and replacing certain tags with 
Server configuration information.  Most of the pages to which 
the Server homepage links directly are built by \script{make-all.pl} 
along with their associated CGI scripts as necessary (for example, the 
\script{list.cgi} script to add 
people to the \emph{neos-news} mailing list).  
Completed HTML files are written under the parent Web directory 
supplied by the Server administrator, 
 and completed CGI scripts are written to the CGI directory
supplied to \script{config.pl}.  

The \script{make-all.pl} script then calls other \script{make} scripts 
to build the more complicated remainder of the Server Web site.  
The \script{make-server-homepage.pl} script builds the Server 
homepage from two files under \file{server/lib/html/}.
The  \script{make-server-solvers.pl} script creates the page listing 
available solvers by reading in registered solvers from the
\file{\var{server\_var}/lib/solver\_list} and \file{server/lib/admin\_list}.
The general Server comments Web page and CGI script are created
by \script{make-server-comments.pl}.
Then \script{make-all.pl} calls on \script{make-solver.pl} to build
Web homes for solvers that the Server already has registered.  
Solver pages are also 
created and updated when solvers are added or modified by the 
\solver{admin:addsolver}, which calls on \script{make-solver.pl} with 
the type and identifier of the desired solver as argument.
In this manner, the Web site stays current without need for the
intervention of the Server administrator.

\pagebreak\section{The NEOS Server FAQ}
\label{faq}
Included in the NEOS Server 4.0 distribution under
\file{server/lib/html/neos\_faq.html} is a
brief  list of debugging-style questions and answers
for solver administrators.
Many of the points in that FAQ are covered here,
but we include them in the FAQ for handy reference.
Also, the FAQ contains much more detailed help with responding to
the Server configuration prompts when you first \command{make}
your NEOS Server.
We would like to hear your own suggestions 
and questions so that we can add them to our 
NEOS Server FAQ at 
\url{http://www-neos.mcs.anl.gov/neos/neos_faq.html}.
Because we may not maintain our NEOS Server at the
same site forever,
you can maintain your FAQ for future administrators
of your own NEOS Server.  

\section*{Acknowledgments}
Work on the NEOS Server first began in 1994 and expanded through the 
collaborative efforts of Joe Czyzyk, Bill Gropp, Mike Mesnier, 
Jorge Mor\'e, Steve Wright, and others.  I especially point out
that Mike Mesnier, as the first NEOS Server administrator, left
me much of the information included in this guide when I 
took over administration and development of the NEOS Server 
for Optimization in 1999.

\addcontentsline{toc}{section}{Acknowledgments}
\end{document}